\documentclass{aa}
\usepackage[english]{babel}
\usepackage{natbib,twoopt}
\usepackage[breaklinks=true]{hyperref}
\bibpunct{(}{)}{;}{a}{}{,}  
\usepackage[utf8]{inputenc}
\usepackage[varg]{txfonts}
\usepackage{adjustbox} 
\usepackage{xspace}
\usepackage{threeparttable}
\usepackage{tikz}
\usepackage{orcidlink}
\usepackage{calc}
\usepackage{xcolor}
\usepackage{amsmath}
\usetikzlibrary{calc} 
\usepackage{booktabs} 
\usepackage{graphicx}
\usepackage{lipsum}
\usetikzlibrary{positioning}  
\usetikzlibrary{arrows.meta}   
\usetikzlibrary{shapes}        
\usetikzlibrary{shapes.multipart} 
\usetikzlibrary{decorations.pathreplacing} 
\usetikzlibrary{positioning, shapes.geometric}

\newcommand{\gsblimLBT}{$31.4$}
\newcommand{\rsblimLBT}{$30.5$}
\newcommand{\Redge}{205$\arcsec$ $\pm$ 5$\arcsec$}
\newcommand{\Rpsf}{190$\arcsec$}
\newcommand{\Sigmaredge}{1.3 $\pm$ 0.1}
\newcommand{\mugedge}{25.95 $\pm$ 0.01 mag/arcsec$^2$}
\newcommand{\muredge}{25.60 $\pm$ 0.01 mag/arcsec$^2$}

\begin{document} 

   \title{LIGHTS. A robust technique to identify galaxy edges}
   
   \author{Giulia Golini \orcidlink{0009-0001-2377-272X} \inst{1,2}, 
   Ignacio Trujillo \orcidlink{0000-0001-8647-2874} \inst{1,2},
   Dennis Zaritsky \orcidlink{0000-0002-5177-727X} \inst{3}, 
   Mireia Montes \orcidlink{0000-0001-7847-0393} \inst{4} ,
   Raúl Infante-Sainz \orcidlink{0000-0002-6220-7133} \inst{5},  
   Garreth Martin \orcidlink{0000-0003-2939-8668} \inst{6},
   Nushkia Chamba \orcidlink{0000-0002-1598-5995} \inst{7},
   Ignacio Ruiz Cejudo \orcidlink{0009-0003-6502-7714} \inst{1,2},
   Andrés Asensio Ramos \orcidlink{0000-0002-1248-0553} \inst{1,2}
   Chen-Yu Chuang \inst{4}, 
   Mauro D'Onofrio \orcidlink{0000-0001-6441-9044} \inst{8}, 
   Sepideh Eskandarlou \orcidlink{0000-0002-6672-1199} \inst{5}, 
   S. Zahra Hosseini-ShahiSavandi \orcidlink{0000-0003-3449-2288} \inst{8},
   Ouldouz Kaboud \orcidlink{0009-0007-7712-0683} \inst{8}, 
   Carlos Marrero de la Rosa \orcidlink{0009-0005-1728-8076} \inst{1,2},
   Minh Ngoc Le \orcidlink{0009-0003-0674-9813} \inst{1,2}, 
   Samane Raji \orcidlink{0000-0001-9000-5507} \inst{9},  
   Javier Román \orcidlink{0000-0002-3849-3467} \inst{10},
   Nafise Sedighi \orcidlink{0009-0001-9574-8585} \inst{1,2},
   Zahra Sharbaf \orcidlink{0009-0004-5054-5946} \inst{1,2}, 
   Richard Donnerstein \orcidlink{0000-0001-7618-8212}\inst{3}, 
   Sergio Guerra Arencibia \orcidlink{0009-0001-7407-2491} \inst{1,2}}

   \institute{Instituto de Astrofísica de Canarias,c/ Vía Láctea s/n, E38205 - La Laguna, Tenerife, Spain 
   \and Departamento de Astrofísica, Universidad de La Laguna, E-38205 - La Laguna, Tenerife, Spain 
   \and Steward Observatory and Department of Astronomy, University of Arizona, 933 N. Cherry Ave., Tucson, AZ 85721, USA 
   \and Institute of Space Sciences (ICE, CSIC), Campus UAB, Carrer de Can Magrans, s/n, 08193 Barcelona, Spain. 
   \and Centro de Estudios de Física del Cosmos de Aragon (CEFCA), Plaza San Juan, 1, E-44001, Teruel, Spain
   \and School of Physics and Astronomy, University of Nottingham, University Park, Nottingham NG7 2RD, UK
   \and NASA Ames Research Center, Space Science and Astrobiology Division M.S. 245-6, Moffett Field, CA 94035, USA 
   \and Department of Physics and Astronomy, University of Padova, Vicolo Osservatorio 3, I-35122, Italy.  
   \and Departamento de Física Teórica, Atómica y Óptica, Universidad de Valladolid, 47011, Valladolid, Spain
   \and Departamento de Física de la Tierra y Astrofísica, Universidad Complutense de Madrid, E-28040 Madrid, Spain 
    }

   \date{}
   \abstract 
   {The LIGHTS survey is imaging galaxies at a depth and spatial resolution comparable to what the Legacy Survey of Space and Time (LSST) will produce in 10 years (i.e., $\sim$31 mag/arcsec$^2$; 3$\sigma$ in areas equivalent to 10$\arcsec$$\times$ 10$\arcsec$). This opens up the possibility of probing the edges of galaxies, as the farthest location of in situ star formation, with a precision that we have been unable to achieve in the past. Traditionally, galaxy edges have been analyzed in one dimension through ellipse averaging or visual inspection. Our approach allows for a two-dimensional exploration of galaxy edges, which is crucial for understanding deviations from disk symmetry and the environmental effects on galaxy growth. 
   In this paper, we propose a novel method using the second derivative of the surface mass density map of a galaxy to determine its edges.  This offers a robust quantitative alternative to traditional edge-detection methods when deep imaging is available. Our technique incorporates Wiener-Hunt deconvolution to remove the effect of the point spread function from the galaxy itself. By applying our methodology to the LIGHTS galaxy NGC 3486, we identify the edge at \Redge{}. At this radius, the stellar surface mass density is $\sim$1 M$_\odot$/pc$^2$, supporting a potential connection between galaxy edges and a threshold for in situ star formation. Our two-dimensional analysis of NGC 3486 reveals an edge asymmetry of $\sim$5$\%$.
   These techniques will be of paramount importance for a physically motivated determination of the sizes of galaxies in ultra-deep surveys such as LSST, Euclid, and Roman.
   }

   
   \titlerunning{Detection of edges of galaxies}
   \authorrunning{Golini et al.}

   \maketitle

\section{Introduction}
\label{sec:intoduction}

In the framework of the $\Lambda$ cold dark matter ($\Lambda$CDM) model, galaxies form and evolve within dark matter (DM) halos. Their mass and size grow over cosmic time through both in situ star formation, whereby gas within the galaxy is converted into stars, and ex situ mechanisms, such as the accretion of stars and gas through mergers or interactions with neighboring systems \citep{toomore, 1978Whiterees, Efstathiou}. While mergers play a significant role in the growth of massive galaxies \citep{Bundy_2009}, for most spirals and dwarf galaxies (see e.g., \citealt{2021garreth}) the primary driver of stellar mass assembly is the transformation of gas into stars, regulated by a critical gas density threshold \citep{1972Quirk, 1989Kennicutt, 2004Schaye, 2008rokroskar}.

The existence of a gas density threshold should leave an imprint on a galaxy’s stellar mass radial density profile. Once the gas reservoir falls below a critical level, star formation ceases, leading to a characteristic “drop” in the stellar mass density. The sharpness of this edge should provide an insight into the recent star formation history of a galaxy. In fact, over time, stellar migration should smooth out these edges \citep{2008rokroskar,2014Sellwood, 2017Debattista}, gradually erasing signatures of past star formation episodes. The prominence of a galaxy’s outer edge would eventually serve as a timer, with sharper edges indicating more recent star formation episodes \citep{Lombilla_2018}.

The critical gas density for star formation is theoretically estimated in the range of 3–10 $M_{\odot}$/pc$^2$ at z = 0 \citep{2004Schaye}.
Accounting for a star formation efficiency of 10–30$\%$, one would expect the stellar mass drop to be found at around 1–3 $M_{\odot}$/pc$^2$. This is in agreement with observational findings for present-day Milky Way-like galaxies \citep{Lombilla_2018, 2022Diazgarcia}.

Detecting systematically such low stellar mass densities require deep imaging with surface brightness limiting depths fainter than $\mu_{g} \sim 26$–27 mag/arcsec$^2$, which is at the detection limit of the Sloan Digital Sky Survey (SDSS; \citealt{2018sdss}). For that reason,
a significant advancement in understanding the gas density threshold for star formation in disk galaxies was made using a low-surface-brightness (LSB) friendly reduction of Stripe 82 data \citep{trujillofliri2016, 2018RomanTrujillo}, which is 2 magnitudes deeper than SDSS. 
\cite{2022chamba-tracinglimits} revealed a moderate mass dependence, with the drop in the stellar mass profile ranging from 0.6 $M_{\odot}$/pc$^2$  in dwarf galaxies to 3 $M_{\odot}$/pc$^2$ in ellipticals.
Using the location of the 1 $M_{\odot}$/pc$^2$ isomass contour (R$_1$; \citealt{2020Trujillo}) as a proxy for the star formation threshold, or the location of the stellar mass density drop as a measurement of the size of the galaxy (R$_{edge}$; \citealt{2022chamba-tracinglimits}), \cite{2020Trujillo} and \cite{2022chamba-tracinglimits} found that the observed scatter in the stellar mass–size relation decreases by a factor larger than 2 with respect to using the effective radius, $R_e$. For galaxies in the mass range $10^7 - 10^{11} M_{\odot}$, the stellar mass–size relation follows a power law with a slope close to $1/3$.
In addition, using this feature as a size measurement, \cite{Buitragotrujillo2024} found an increase in the size of the disk by a factor of 2, since z $\sim 1$ for Milky Way-like disk galaxies (M$_{stellar} \sim  5 \times 10^{10} M_{\odot}$).\\
The vast majority of photometric studies analyzing breaks and/or galaxy edges rely on one-dimensional (1D) surface brightness or stellar mass density profiles \citep{2006pohlen, Bakos2008, 2011Gutierrez, 2012COMERON, 2012martinnavarro, 2017Peterswaywemeasure, 2017peters, 2020Trujillo, 2022chamba-tracinglimits, 2024Chambafornax, Buitragotrujillo2024}. While effective, these methods average over azimuthal variations and substructures, potentially missing key morphological features. In contrast, bidimensional (2D) analyses preserve spatial information, allowing to capture asymmetries and localized angular substructures.
However, such analyses require deeper data ($\mu_{g} >$ 28 mag/arcsec$^2$) to be feasible and reliable. \\ 
Optimized observational strategies and data reduction methods \citep{2010Jablonka, trujillofliri2016, LIGHTSs, montes2021buildup, 2023romancoma, 2024montesnube, 2024Golini} now allow for surface brightness limits of $\mu_{g} > 30$ mag/arcsec$^2$ (3$\sigma$; 10$\arcsec$ $\times$ 10$\arcsec$  boxes) and similar depths are expected from the upcoming 10-year Legacy Survey of Space and Time (LSST; \citealt{2019lsst}). These deeper data are crucial for systematically investigating the edges of galaxies with an unprecedented accuracy.\\
Therefore, with the increasing volume of deep observational data, a robust, automated, and unbiased method of identifying galaxy edges is essential, as previous studies have primarily relied on visual detection.
In computer vision, boundaries are identified by detecting rapid transitions in brightness, color, or texture (see \citealt{JINGrewievedges} for a review). However, defining a galaxy's edge requires a physically motivated approach rather than purely visual cues. While some methods, such as fitting intersecting exponential functions \citep{2023lombilla}, have been proposed to objectively locate edges, a fully automated approach is crucial for ensuring consistency across large datasets.
A recent study by \cite{2024jesusML} applied artificial intelligence to automate the identification of galaxy edges using Hubble images. By analyzing different combinations of color images, their algorithm successfully recovered previously visually identified edges \citep{Buitragotrujillo2024}. Similarly, \cite{Vega-Ferrero2025} applied the segment anything model to recover the radial positions of the stellar disk edges from \cite{Buitragotrujillo2024}.
However, machine learning models are often difficult to interpret.
The goal of our work is to present a robust method of identifying galaxy edges by locating the minimum of the second derivative of the stellar mass density distribution. This technique provides a reproducible approach to measuring galaxy sizes across large datasets, facilitating a more systematic study of their outermost regions. Additionally, it will enable the investigation of the stellar mass density threshold at the edge across different morphologies, environments \citep{2024Chambafornax}, and will allow for the exploration of the cosmic evolution of galaxy sizes \citep{Buitragotrujillo2024} with the upcoming next generation of datasets.\\
This paper is structured as follows. Section \ref{sec:Amotivateddefinition} describes the methodology to locate the edge of galaxies. Section \ref{sec:data} presents the LBT Imaging of Galaxy Haloes and Tidal Structures (LIGHTS; \citealt{LIGHTSs, 2024Zaritskydennislights}) data used in this work for the galaxy NGC 3486. Section \ref{sec:analysis} details data processing and mitigation of scattered light effects. Sections \ref{sec:discussion} and \ref{sec:conclusions} discuss our findings and summarize conclusions. All magnitudes are reported in the AB system \citep{okegunn}.

\section{A robust procedure to locate galaxy edges}
\label{sec:Amotivateddefinition}

We propose a method that uses the second derivative of the stellar mass density distribution, $\Sigma_{*}(r)$, to determine a galaxy's edge. By finding the radial position at which the second derivative reaches the outermost\footnote{The term “outermost” is used because inner regions of a galaxy, such as the bulge, bar transition, spiral arms, or clumps, can also produce additional minima.} minimum, we identify the point of  curvature corresponding to the outer sharp change in slope in $\Sigma_{*}(r)$. This point marks a natural transition in the galaxy’s structure and provides a robust identification of its edge.

\subsection{Motivation for using the stellar mass density profile}
\label{motivationstellardensity}

If a gas density threshold is indeed present \citep{2004Schaye, 2008rokroskar} affecting star formation, both surface brightness and stellar mass density profiles should reflect a structural transition: one in terms of light ($\mu$) and the other in terms of mass ($\Sigma_{*}$).
In this work we use the stellar mass density profile.
The reason lies in the nature of the stellar populations. In surface brightness profiles, particularly in bluer bands, variations in the luminosity of young star-forming regions, spiral arms, and clumps introduce localized fluctuations that can create multiple substructures in the profile. These can blur the detection of the outermost drop in the averaged lights profile, as they introduce local minima of the second derivative. 
In contrast, the stellar mass density profile, derived from the full spectral energy distribution or from color combinations, smooths out variations in stellar populations across the disk, reducing the influence of substructures that are more prominent in light-based profiles (see Fig. \ref{fig:all-sbl}).
Working with $\Sigma_{*}(r)$ profiles also has some drawbacks. The most important one is the addition of larger uncertainties as it relies on the combination of multiple surface brightness profiles plus the adoption of some stellar population models. For these reasons, the ideal scenario would be to work with ultra-deep near-infrared data, for which the underlying mass distributions can be traced more directly.
In this work we rely on $g$ Sloan and $r$ Sloan data to get the stellar mass density profile of galaxies, as these are the available filters in the LIGHTS survey.

\subsection{Mathematical derivation}
\label{edge-definition}

As was mentioned in the introduction, the end of in situ star formation should lead to a drop in the stellar mass distribution of a galaxy disk.
Therefore, in the majority of galaxy disks and dwarfs, the mass distribution should be well represented by a broken exponential function (a function described in \citealt{2008Erwin}) that has two defined exponential declining zones with different scale lengths. We define the transition point as the edge of the galaxy, $R_\textnormal{edge}$.

Within the galaxy edge (r < $R_\mathrm{edge}$), the slope of the disk’s stellar mass surface density profile, $\Sigma_{*}(r)$, is parametrized by the scale length, $h_1$. Beyond $R_\mathrm{edge}$, the profile declines with a second scale length, $h_2$.
The sharpness of the edge is parametrized by $\alpha$.
For an infinitely thin disk, the broken exponential is described as

\begin{equation}
    \Sigma_{*}(r) \propto e^{-\frac{r}{h_1}} \left[1 + e^{\alpha(r - R_\textnormal{edge})}\right]^{\frac{1}{\alpha} \left(\frac{1}{h_1} - \frac{1}{h_2}\right)}
    \label{general_formula}
.\end{equation}

The first derivative, $\frac{\partial}{\partial r} \Sigma_{*}(r)$,  with respect to the radial position indicates how the intensity changes along the radius. 
For a broken exponential function on a logarithmic scale, $\mathrm{Log}(\Sigma_{*}(r))$, 
the first derivative, $\frac{\partial}{\partial r} \mathrm{Log} (\Sigma_{*}(r))$, shows a sharp change at the point where the slope changes, $R_\mathrm{edge}$, preceded and followed by two constants (equal to - 1/$h_{1}$ before $R_\mathrm{edge}$ and to - 1/$h_{2}$ beyond). 
The second derivative, $\frac{\partial^2}{\partial r^2} \mathrm{Log}(\Sigma_{*}(r))$, gives information about the curvature or the rate of change in the slope. At the point where the slope changes abruptly, the second derivative has a minimum.
Therefore, $\frac{\partial^2}{ \partial r^2} \mathrm{Log} (\Sigma_{*}(r))$ presents a minimum corresponding to $R_\mathrm{edge}$.
In other words, the search for the minimum of $\frac{\partial^2}{\partial r^2} \mathrm{Log} (\Sigma_{*}(r))$ corresponds to the search for the location of $R_\mathrm{edge}$.

\subsection{Application to a simulated disk}
\label{imfitdisc}

\begin{figure*}[!ht]
    \centering
    \includegraphics[width=\linewidth]{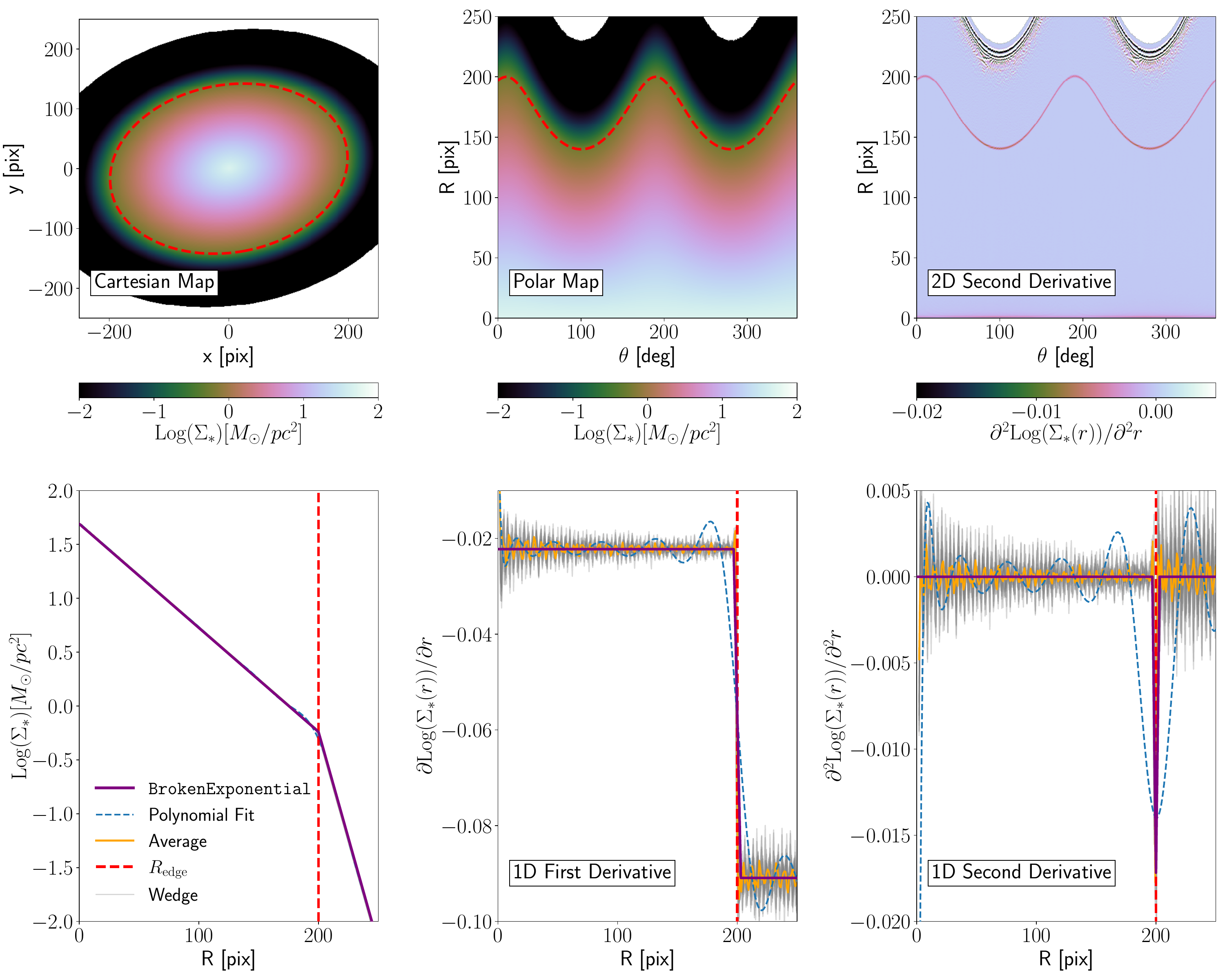}
    \caption{Edge detection method. The figure presents the stellar mass density map in Cartesian and polar coordinates (top left and top central panels), and the averaged surface stellar mass density profile (bottom left panel) of a simulated disk, computed in three different cases as is explained in the main text. The first derivative of the stellar mass density profile (bottom central panel) and the second derivative in both 2D (upper right) and 1D (bottom right) are also shown. Gray lines represent the surface stellar mass density and its derivatives using 10 degrees wedges along azimuthal angles. Dashed red lines indicate \( R_\textnormal{edge} \), corresponding to the minimum of the second derivative.}
    \label{fig:main-imfit}
\end{figure*}

To illustrate the implementation of our second-derivative method of edge detection,
we generated a simulated disk using IMFIT \citep{Erwin2015}, an image-fitting tool designed for modeling galaxy surface brightness profiles. The disk follows a \texttt{BrokenExponential} light distribution, with an inner scale length of $h_{1}$ = 45 pixels transitioning to a steeper declining outer slope of $h_2$ = 11 pixels at $R_\mathrm{edge}$ = 200 pixels. We assume a sharp edge with $\alpha$ = 10, a position angle (PA) of 30 deg (counterclockwise from the X axis), and an axis ratio (AR) of 0.7.

The top left panel of Fig. \ref{fig:main-imfit} shows the stellar mass density distribution of the simulated disk in Cartesian coordinates, where the dashed ellipse marks the point of change in the slope ($R_\mathrm{edge}$). The top central panel presents the same distribution in polar coordinates and the sinusoidal wave corresponding to the edge location, which allows one to examine possible asymmetries in the edge structure.

\subsubsection{1D edge detection}
\label{1dexplanation}

The bottom row of Fig. \ref{fig:main-imfit} summarizes the 1D analysis, showing (from left to right) the stellar mass density profile, its first derivative, and its second derivative in three different cases: the analytical \texttt{BrokenExponential} function used to generate the simulated disk with IMFIT, the averaged stellar mass surface density profile of the disk, and the 30-degree\footnote{A high-degree polynomial capture subtle variations and provides high fitting precision.} polynomial that fits the averaged stellar mass surface density profile of the disk.
The averaged stellar mass surface density profile of the disk was extracted using elliptical apertures with a fixed PA and inclination, and a radial step size of one pixel with the \textit{astscript-radial-profile} in Gnuastro \citep{astscriptradialprofileRaul}.

The derivatives of the \texttt{BrokenExponential} function and the polynomial fit were calculated analytically. The derivatives
of the averaged stellar mass density profile of the simulated disk were computed using a custom implementation of the \texttt{five-point-stencil} derivative\footnote{The error in this approximation of derivatives is of order 4 \citep{abramowitz1948handbook}.} method in Python \citep{SAUER}. This numerical method leverages a higher-order Taylor series expansion, yielding a more precise derivative by drawing on multiple neighboring points. The approach reduces local errors and smooths out irregularities, though it does cause the loss of the first and last data points.
However, this is inconsequential here, as our primary focus is on the outer regions of the galaxy, which are well covered within our data range.
The edge location remains consistent in the three cases, demonstrating the robustness of our method using simulated data.
The minimum of the second derivative corresponds to $R_\mathrm{edge}$ (dashed red lines in Fig. \ref{fig:main-imfit}).

\subsubsection{2D edge detection}
\label{2dexplanation}

While fixed elliptical apertures are a standard method in surface photometry, they average over the entire azimuthal range, potentially smoothing out localized variations. In contrast, wedge\footnote{A wedge is a sector of a circle that is defined by two straight lines (radii) extending from the center of the outer regions, and the arc connecting them.}-based profiles allow for a more localized analysis, capturing directional differences and small-scale features, even though they still require assumptions about the geometry in each sector.

The elliptical symmetry of the simulated disk ensures that the minimum of the second derivative along each radial direction, $R_{\textnormal{edge},\theta}$, corresponds to $R_\mathrm{edge}$ by design.
The top right panel of Fig. \ref{fig:main-imfit} presents the second derivative map in polar coordinates\footnote{In this image, each column is treated as a 1D array, and the pixel values represent the numerical second derivatives computed along each column.}, where a clear sinusoidal pattern emerges (in pink), corresponding to the edge. This structure marks the edge in different azimuthal directions. 
The dispersion between gray lines in the bottom panels of Fig. \ref{fig:main-imfit} represents the uncertainty in computing the derivatives using 10 degrees wedges along different azimuthal angles.
No error bars are shown, as we assume an idealized scenario with infinite signal to noise (S/N).
Despite the disk model being noise-free, minor inaccuracies in locating R$_{\textnormal{edge},\theta}$ and artificial oscillations of the derivatives arise from numerical artifacts introduced during the discrete generation of the 2D image, and radial or angular binning of the profile.
The choice of width of the wedge, depending on the S/N of real data, is discussed in the next sections.

\section{Target selection and data}
\label{sec:data}

We applied the methodology described above to the LIGHTS data of NGC 3486, an M33-like galaxy (i.e., with a similar stellar and total mass to M33) located at 13.6 Mpc \citep{kourkchi2020cosmicflows} and in a region of low Galactic extinction (A$_g$ = 0.08, A$_r$ = 0.06;
\citealt{2011schlafly}).
At this distance, the full width at half maximum (FWHM) of point-like sources of LIGHTS data is 1.6$\arcsec$ corresponding to 100 pc.
NGC 3486 constitutes a case study; we aim to perform the analysis to the entire LIGHTS dataset. 
NGC 3486 was observed with the Large Binocular Cameras (LBCs) Blue and Red mounted at the Large Binocular Telescope (LBT) in Arizona as part of the program DDT2019B12 (PI: M. D’Onofrio). 
Each camera comprises four CCDs with a pixel scale of 0.224 $\arcsec$/ pix.
The LBC setup includes four CCDs, with each CCD covering 7.8$\arcmin$ $\times$ 17.6$\arcmin$ section of the sky and a total camera field of view (FoV) of $\sim$ 23$\arcmin$ $\times$ 25$\arcmin$.
For these observations, the $g$ Sloan filter was employed in the LBC Blue camera, and the $r$ Sloan filter was utilized in the LBC Red camera.
The data collection took place on April 3, 2021, under dark sky conditions. By using dithered exposures of 180 seconds each, with four arcminutes offsets, we obtained a total on-source integration time of 1.5 hours (see \citealt{2024Zaritskydennislights} for more details on observations).\\
The data reduction process followed a similar strategy to the one outlined in \cite{LIGHTSs} and developed in \cite{2024Zaritskydennislights}.
Our pipeline’s main steps involved deriving flat field images, background removal, astrometric and photometric calibrations, and mosaic co-addition.
The final stacked data cover a region of 40.2$\arcmin$ $\times$ 40.2$\arcmin$ around NGC 3486.
With this observational strategy and optimized data reduction we achieved a surface brightness limiting depth of $\gsblimLBT{}$ mag/arcsec$^2$ and $\rsblimLBT{}$ mag/arcsec$^2$ in the $g$ and $r$ Sloan bands, respectively, corresponding to a 3$\sigma$ fluctuation of the background in areas equivalent to 10$\arcsec$ $\times$ 10$\arcsec$ \citep{trujillofliri2016, roman2020}.
Our data allow us to reach surface brightness limits about 4 mag/arcsec$^2$ deeper than the previous imaging of this particular source with SDSS.
We show in Appendix \ref{app:rgb} the full FoV of the LIGHTS data of NGC 3486.

\section{Methods}
\label{sec:analysis}

The edge detection process for NGC 3486 follows a structured approach. We began by masking foreground and background sources to minimize contamination. Next, although LIGHTS data are background-subtracted, in some cases the subtraction was not precise enough to reliably represent the local surrounding background of the galaxy, leading to slight over- or underestimation. Therefore, we corrected for the local background. Then, we applied the point spread function (PSF) correction to account for the scattering of the galaxy's light due to the effect of the PSF. After this, we proceeded to get the surface stellar mass density map of NGC 3486. Finally, we derived the location of the edge of NGC 3486 using the methodology described in Sect. \ref{sec:Amotivateddefinition}.
Figure \ref{fig:deconvolutionstep} provides a schematic overview that serves as a reference for the methodology presented in the following sections.

\newcommand\ImageNode[3][]{
  \node[draw=black, line width = 2pt, #1] (#2) {\includegraphics[width=4.cm, height=4.cm]{#3}};}

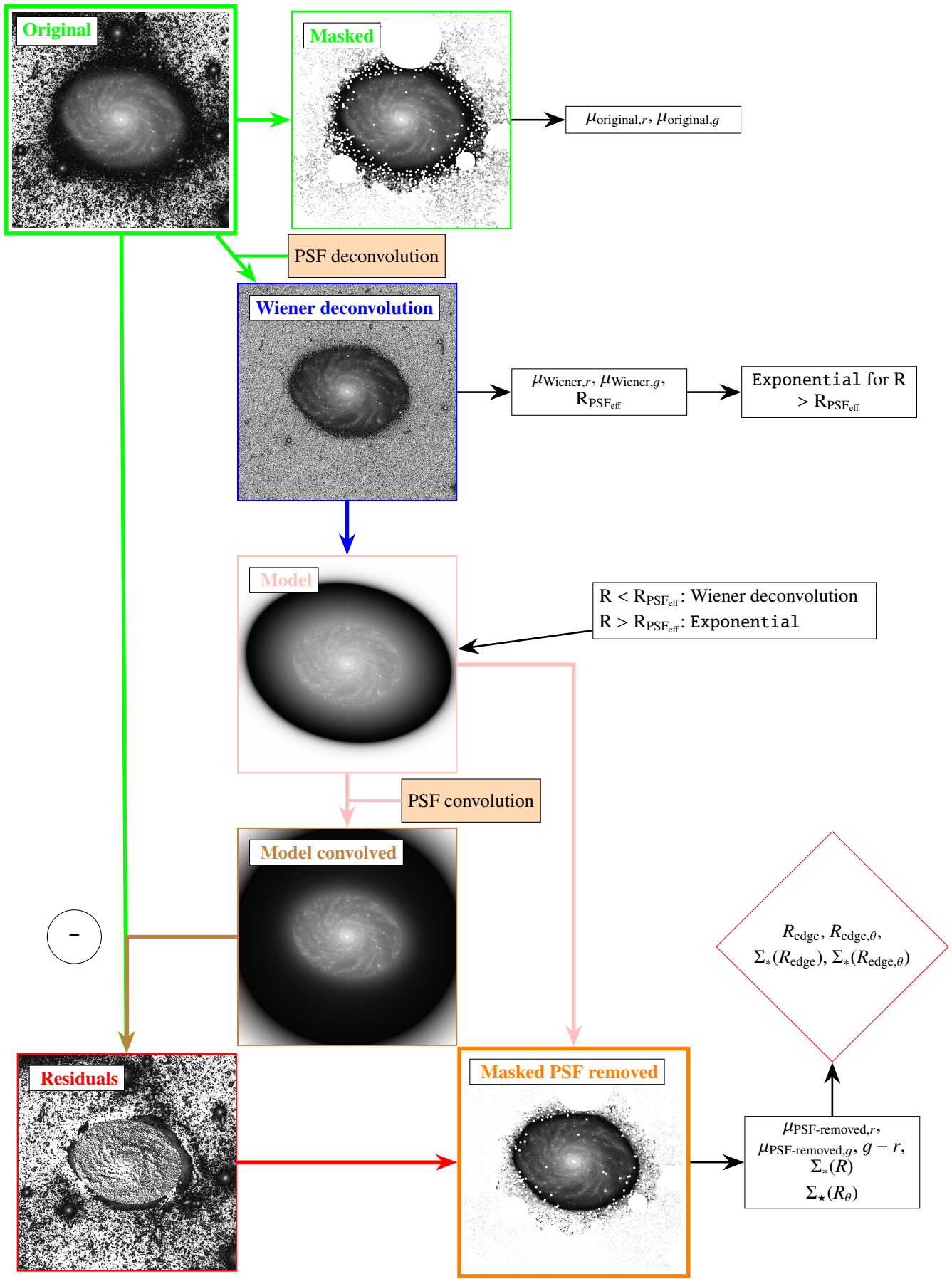
\begin{figure*}
    \centering   

        \tikzstyle{decision} = [rectangle, minimum width=2cm, minimum height=0.5cm, text centered, draw=black, fill=white]
        \tikzstyle{decisiondef} = [diamond, minimum width=2cm, minimum height=0.5cm, text centered, draw=red, fill=white]
        
        \tikzstyle{process} = [rectangle, minimum width=1.3cm, minimum height=0.8cm, text centered, draw=black, fill=orange!30]

        \tikzstyle{parallel} = [parallelogram, minimum width=3cm, minimum height=1cm, text centered, draw=black, fill=white]
        
        \begin{tikzpicture}[node distance=1cm]
        
            \ImageNode[draw = green]{A}{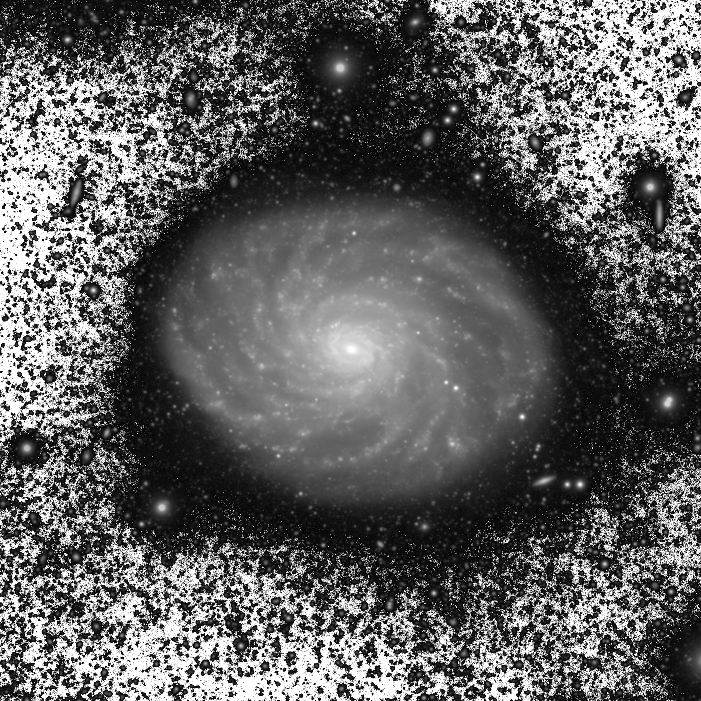}
            \node[anchor=north west, xshift=0.1cm, yshift=-0.1cm] at (A.north west)  {\fcolorbox{black}{white}{\textbf{\textcolor{green}{Original}}}};

            \ImageNode[draw = green, label={}, inner sep=0pt, right= of A]{B}{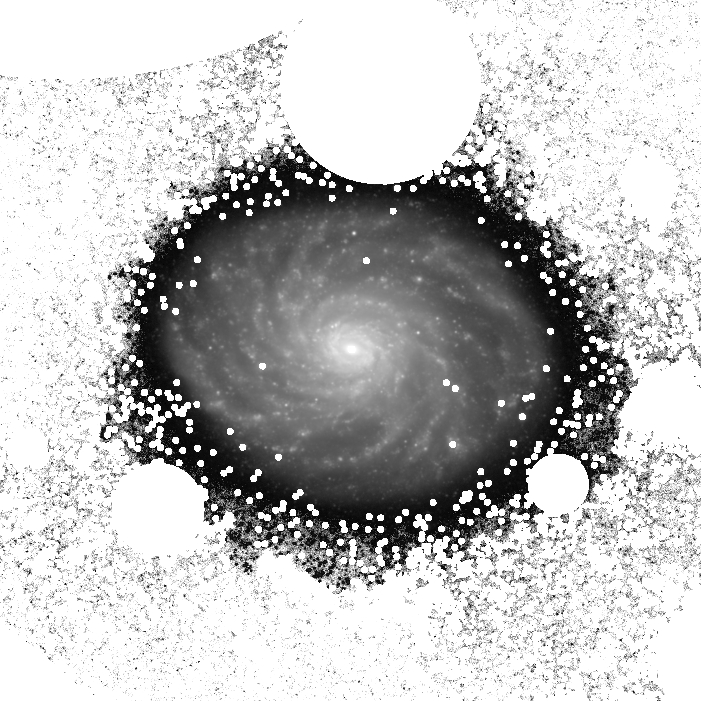}
            \node[anchor=north west, xshift=0.1cm, yshift=-0.1cm] at (B.north west)  {\fcolorbox{black}{white}{\textbf{\textcolor{green}{Masked}}}};
            
            \draw[->, thick, color=green, line width=2pt, >=Stealth, >={Stealth[length=4mm, width=3mm]}](A) -- (B);

            \node (profs1) [decision, right = of B] 
            {\parbox{3cm}{\centering 
            $\mu_{\textnormal{original},r}$, $\mu_{\textnormal{original},g}$}};

            \draw[->, thick, color=black, line width=1pt, >=Stealth, >={Stealth[length=4mm, width=3mm]}](B) -- (profs1);

            \ImageNode[draw = blue, label={}, inner sep=0pt, below= of B,xshift=-1cm]{C}{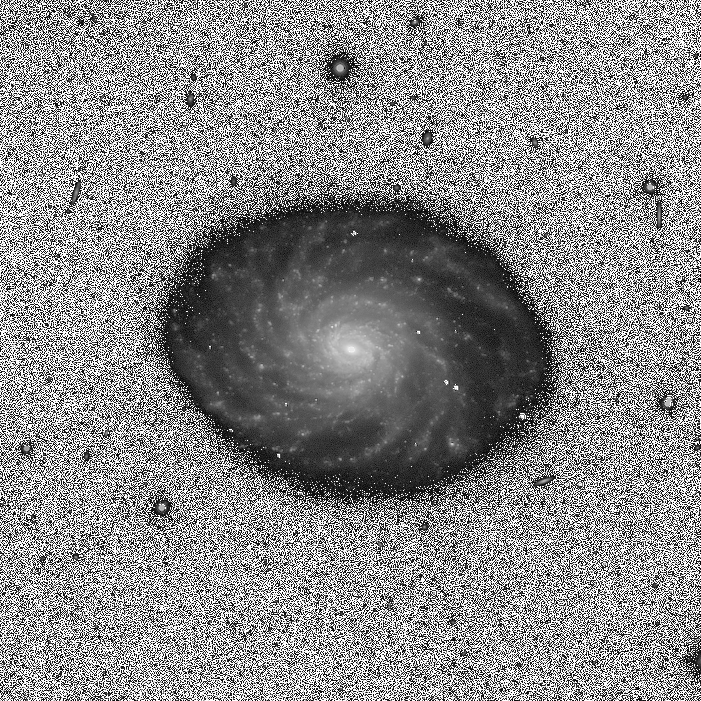}
            \node[anchor=north west, xshift=0.1cm, yshift=-0.1cm] at (C.north west) 
            {\fcolorbox{black}{white}{\textbf{\textcolor{blue}{Wiener deconvolution}}}};

            \draw[->, thick, color=green, line width=2pt, >=Stealth, >={Stealth[length=4mm, width=3mm]}](A) -- (C);
            
            \node (pro1) [process, right = of  $(A)!0.5!(C)$] {PSF deconvolution};
            \draw[-, thick, color=green, line width=1.5pt](pro1) -- ($(A)!0.5!(C)$);

            \node (profs3) [decision, right = of C] 
            {\parbox{3cm}{\centering 
            $\mu_{\textnormal{Wiener},r}$, $\mu_{\textnormal{Wiener},g}$, R$_{\textnormal{PSF}_\textnormal{eff}}$}};

            \node (profs4) [decision, right = of profs3] 
            {\parbox{3cm}{\centering \texttt{Exponential} for R > R$_{\textnormal{PSF}_\textnormal{eff}}$}};
            
            \draw[->, thick, color=black, line width=1pt, >=Stealth, >={Stealth[length=4mm, width=3mm]}](profs3) -- (profs4);

            \draw[->, thick, color=black, line width=1pt, >=Stealth, >={Stealth[length=4mm, width=3mm]}](C) -- (profs3);

            \ImageNode[draw = pink, label={}, inner sep=0pt, below = of C]{D}{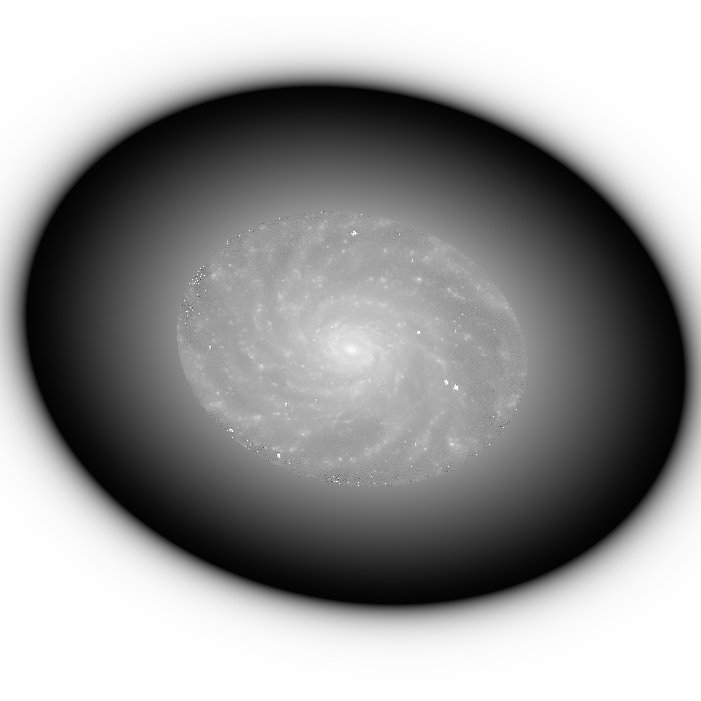}

            \node (paranode) [rectangle, minimum width=3cm, minimum height=0.8cm, text centered, draw=black, fill=white, right = of D, xshift=1.5cm,yshift=1cm] {\parbox{5cm}{ R < R$_{\textnormal{PSF}_\textnormal{eff}}$:  Wiener deconvolution \\[0.1cm] R > R$_{\textnormal{PSF}_\textnormal{eff}}$: \texttt{Exponential}}};

            \draw[->, thick, color=black, line width=1pt, >=Stealth, >={Stealth[length=4mm, width=3mm]}](paranode) -- (D);
            
            \node[anchor=north west, xshift=0.1cm, yshift=-0.1cm] at (D.north west) {\fcolorbox{black}{white}{ \textbf{\textcolor{pink}{Model}}}};

            \draw[->, thick, color=blue, line width=2pt, >=Stealth, >={Stealth[length=4mm, width=3mm]}](C) -- (D);

            \ImageNode[draw = brown, label={},inner sep=0pt, below = of D]{E}{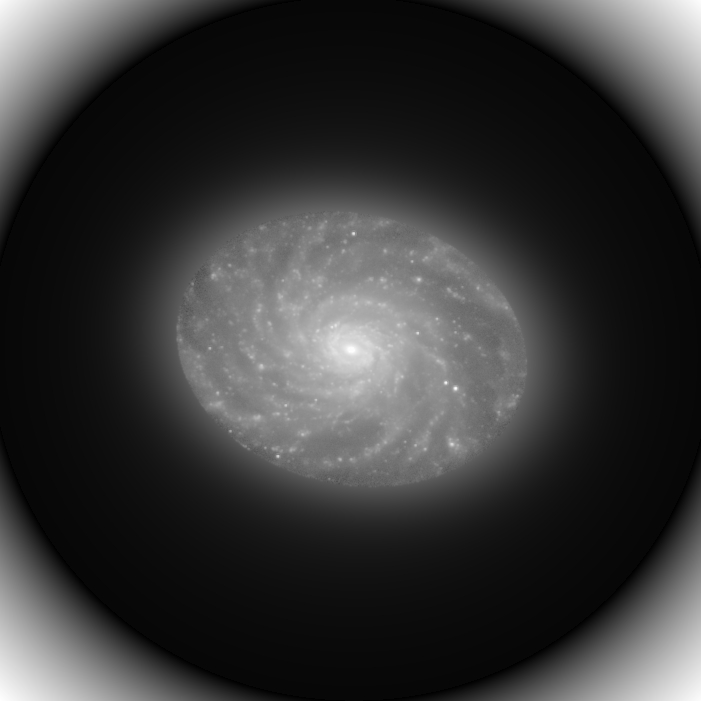}

            \node (minusnode) [circle, minimum size=1cm, text centered, draw=black, fill=white, left = of E,xshift=-1.5cm] {\textbf{--}};
            
            \node[anchor=north west, xshift=0.1cm, yshift=-0.1cm] at (E.north west) {\fcolorbox{black}{white}{ \textbf{\textcolor{brown}{Model convolved}}}};

            \draw[->, thick, color=pink, line width=2pt, >=Stealth, >={Stealth[length=4mm, width=3mm]}](D) -- (E);

            \ImageNode[draw = red,label={}, inner sep=0pt, left = of E, yshift=-4.2cm,xshift=1cm]{F}{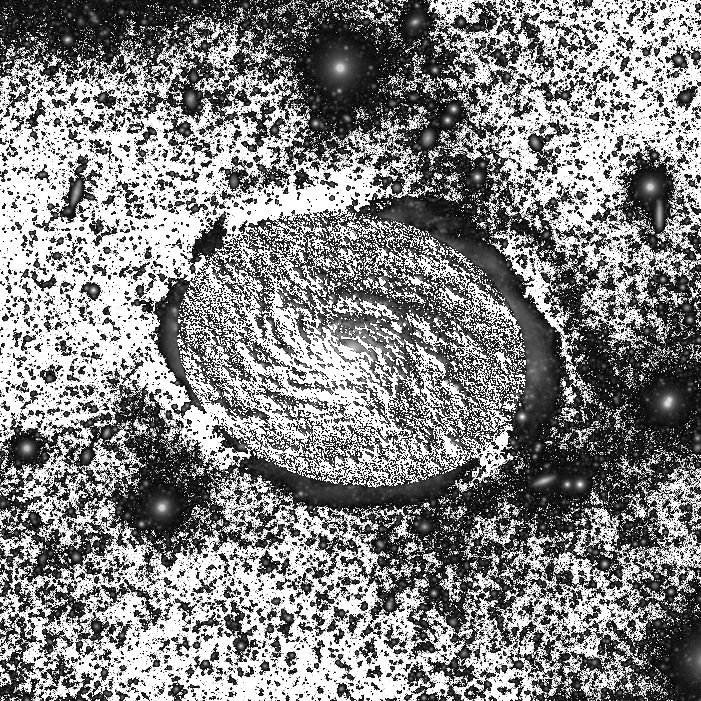}
            
            \node[anchor=north west, xshift=0.1cm, yshift=-0.1cm] at (F.north west) {\fcolorbox{black}{white}{ \textbf{\textcolor{red}{Residuals}}}};

            \ImageNode[draw = orange,label={}, right=of E,yshift=-4.2cm,xshift=-1cm]{G}{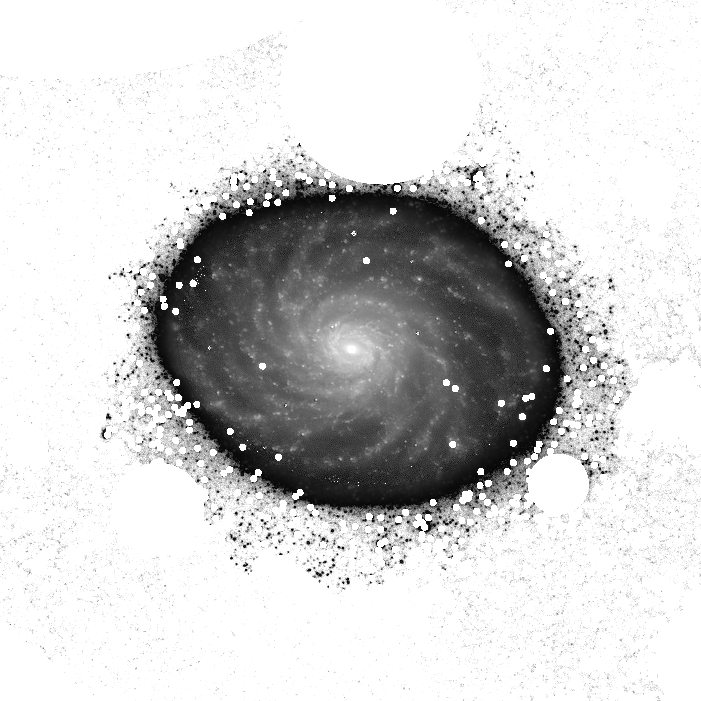}
            
            \node[anchor=north west, xshift=0.1cm, yshift=-0.1cm] at (G.north west) {\fcolorbox{black}{white}{ \textbf{\textcolor{orange}{Masked PSF removed}}}};

            \node (profs10) [decision, right = of G] 
            {\parbox{3cm}{\centering 
            $\mu_{\textnormal{PSF-removed},r}$, $\mu_{\textnormal{PSF-removed},g}$, $g-r$, $\Sigma_{*}(R)$\\[0.1cm] $\Sigma_{\star}(R_{\theta})$}};

            \node (profs20) [decisiondef, above = of profs10] 
            {\parbox{3cm}{\centering 
            $R_\mathrm{edge}$, $R_{\textnormal{edge},\theta},$\\[0.1cm] $\Sigma_{*}(R_\mathrm{edge})$,  $\Sigma_{*}(R_{\textnormal{edge},\theta})$}};

            \draw[->, thick, color=black, line width=1pt, >=Stealth, >={Stealth[length=4mm, width=3mm]}](G) -- (profs10);
            
            \draw[->, thick, color=black, line width=1pt, >=Stealth, >={Stealth[length=4mm, width=3mm]}](profs10) -- (profs20);

            \draw[->, thick, color=pink, line width=2pt, >=Stealth, >={Stealth[length=4mm, width=3mm]}](D) -| (G);

            \node (pro2) [process, right = of  $(D)!0.5!(E)$] {PSF convolution};
            
            \draw[->, thick, color=green, line width=2pt, >=Stealth, >={Stealth[length=4mm, width=3mm]}](A) -- (F);
            \draw[->, thick, color=brown, line width=2pt, >=Stealth, >={Stealth[length=4mm, width=3mm]}](E) -| (F);

            \draw[->, thick, color=red, line width=2pt, >=Stealth, >={Stealth[length=4mm, width=3mm]}](F) -- (G);

            \draw[-, thick, color=pink, line width=1.5pt](pro2) -- ($(D)!0.5!(E)$);

        \end{tikzpicture}
        
    \caption{Flowchart illustrating the steps used to account for the effect of the PSF on the outermost regions of NGC 3486 (see main text for details). The panels display LIGHTS data of NGC 3486. Each panel covers a FoV of 15.5$\arcmin$ $\times$ 15.5$\arcmin$.}
    \label{fig:deconvolutionstep}

\end{figure*}

\subsection{Mask and local background treatment}
\label{sky-mask}

To achieve accurate background subtraction, we first masked foreground and background sources. The mask was generated by combining the $g$ and $r$ band masks. The first step in creating the masks involves running \texttt{NoiseChisel} and \texttt{Segment} \citep{gnuastro, noisechisel_segment_2019}. To improve the detection of contaminating sources, we applied unsharp masking, which removes the light from the galaxy and facilitates the identification of individual contamination sources. Unsharp masking involves smoothing the original image with a Gaussian filter corresponding to the FWHM of point-like sources and subtracting this smoothed image from the original. Finally, we manually inspected the deep color image of NGC 3486 (see Fig. \ref{fig:ngc3486rgb}) to mask any remaining undetected objects.
The final mask applied to NGC 3486 is shown in Fig. \ref{fig:mask} in Appendix \ref{app:maskimage}.

To compute the local background correction, once the data were masked, we manually placed circular apertures (each with a 20$\arcsec$ radius) in specific regions approximately 6.5$\arcmin$ from the galaxy center. These locations were selected to be relatively free of contamination, avoiding bright sources ($V < 17$ mag) and diffuse light brighter than $\mu_g \sim 28$ mag/arcsec$^{2}$. We adopted this conservative approach to ensure robust background estimation. We then computed the resistant (3$\sigma$-clipped) median of the unmasked background pixels within these regions.
This process was performed separately for each filter.

\subsection{Removing the effect of the PSF in the outermost regions of the galaxy}
\label{sec:psfremoval}

Light scattering from the PSF can significantly alter the observed properties of a galaxy’s outer disk by redistributing mostly light from the central regions, leading to an artificial brightening of the outskirts \citep{2001trujillopsf, Sandin2014,2015SANDIN}. The effect depends on the instrument and wavelength used \citep{2015Duc,2006deccroston,2015mihos,2014Abrahamvd}.
In this work, we mitigated PSF effects in the galaxy’s outermost regions by combining model-based convolution techniques \citep{trujillofliri2016,2019Rich, 2017peters, 2019lombilla, infante2020sloan} with direct image deconvolution methods, such as \texttt{PyOperators} \citep{2017karabal}. The following sections detail the steps taken in this process.

\subsubsection{Breaking the PSF-Stellar halo degeneracy}
\label{psf-halodeg}

\begin{figure*}[h!]
    \centering
    \includegraphics[width=\textwidth]{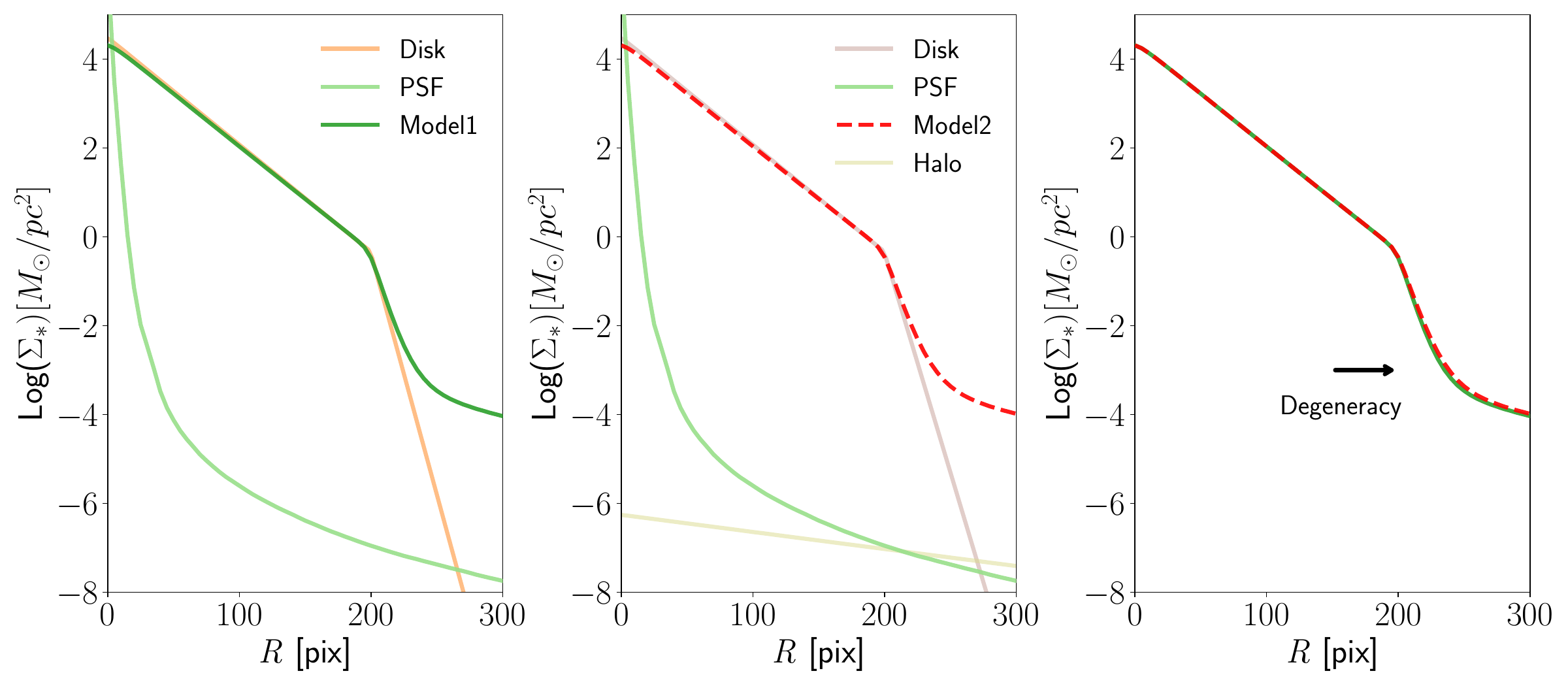}
    \caption{Surface stellar mass density profiles of two simulated disks. Left: Model1, consisting of a broken exponential disk only. Center: Model2, similar to Model1, but with a steeper outer disk slope ($h_2$) and a faint stellar halo. Right: Comparison of the PSF-convolved profiles for both models.}
    \label{fig:degeneracy}
\end{figure*}

\textbf{The problem}:
It can be challenging to distinguish a faint stellar halo present around galaxies from the light-broadening effects caused by the PSF. This confusion, which we term PSF-Stellar halo degeneracy, makes it difficult to tell if a galaxy’s outer light distribution is due to an extended stellar halo or merely an artifact of the PSF. 
To illustrate the problem, we generated two \texttt{BrokenExponential} models using the \texttt{makeimage} task in IMFIT, with the same PA, AR, and outer edge radius, $R_{\textnormal{edge}}$.
Model1 has an outer slope ($h_{2}$) that is 20$\%$ shallower than the one of Model2.
In addition, Model2 is composed of a faint \texttt{Exponential} stellar halo component.
We applied the extended $g$ band PSF of the LBC \citep{2024sedighi} to both models.
The averaged stellar mass surface densities of Model1 and Model2 are illustrated in the left and central panels of Fig. \ref{fig:degeneracy}, respectively, along with their components and the profile of the PSF. 
The parameters used to create the models are reported in Appendix \ref{app:modelsdegeneracy}.

The outer profile (R > 200 pix) in both models can be interpreted as either only a PSF effect on the declining disk light or a faint stellar halo combined with the PSF. This uncertainty highlights the PSF-Stellar halo degeneracy, making it difficult to distinguish between PSF broadening and a different outer profile, $h_{2}$, plus a faint stellar halo.
This test can be extended to higher inclinations, which amplify PSF effects, emphasizing the need for accurate PSF deconvolution when analyzing faint outer structures in galaxies. 

In summary, the PSF redistributes photons, artificially broadening the outer regions of galaxies and smoothing the slope of the outer disk ($h_2$). In general this depends on the shape of the PSF. This redistribution does not alter the location of the edge in the stellar mass density distribution. However, it is key to correct for the PSF to ensure accurate recovery of the intrinsic shape of the profile, particularly the outer slope ($h_2$).\\

\textbf{Wiener-Hunt deconvolution to break the degeneracy:} 
We used the extended PSF of the LIGHTS survey \citep{2024sedighi} to deconvolve the background-subtracted data by applying the Wiener-Hunt \citep{Orieux} deconvolution algorithm.
To mathematically frame the effect of the PSF, the observed data $g(x,y)$ from the telescope can be expressed as a convolution of the “true” image, $f(x,y)$, with the PSF, $h(x,y)$, plus an additive noise term, $\eta(x,y)$:

\begin{equation}
    g(x,y) = f(x,y) * h(x,y) + \eta(x,y)
    \label{eq:image_formation}
.\end{equation}

In an ideal, noise-free scenario, we can use Fourier transforms to work with these data components in frequency space. Specifically, the Fourier transforms, $G(u, v)$ of $g(x, y)$ and $H(u, v)$ of $h(x, y)$, would allow us to isolate $F(u, v)$, the Fourier transform of the true image. From this, we could reconstruct $f(x,y)$ by transforming back into spatial space. However, because $H(u,v)$ acts as a low-pass filter, it tends to amplify high-frequency noise. Thus, a noise suppressor is needed.
The deconvolution process using Wiener-Hunt filter was performed in the Fourier domain:

\begin{equation}
F(u,v) = \frac{G(u,v) H^*(u,v)}{|H(u,v)|^2 + K}
,\end{equation}

where $H^{*}(u,v)$ is the complex conjugate of $H(u,v)$ and $K$ is the regularization parameter that controls noise suppression.
In this work we used Wiener-Hunt deconvolution implemented in the Python package \texttt{skimage}\footnote{ \texttt{skimage.restoration.unsupervised.Wiener}}, in which the regularization parameter is automatically estimated using a stochastic iterative process \citep{Orieux}.

To visualize the effect of performing the Wiener-Hunt deconvolution on the data, we extracted the surface brightness profiles for NGC 3486 in both $g$ and $r$ bands with \textit{astscript-radial-profile} and a bin size of 10$\arcsec$\footnote{Larger bin sizes smooth the profiles by averaging the flux over wider elliptical annuli, thereby reducing point-to-point fluctuations. In contrast, smaller bin sizes increase the sensitivity to small-scale variations, leading to more oscillatory behavior in the derivative profiles. We verified that the choice of step size does not affect the determination of the edge location.}.
We extracted the light profiles of the galaxy using AR = $0.765 \pm 0.020$ and a PA of $13.0 \pm 0.7^\circ$ (clockwise from the X axis). This choice of parameters is justified in the Appendix \ref{stellar-geometry} and compared with previous estimates.
To maintain consistency, we applied the same mask used for sky background estimation (see Fig. \ref{fig:mask}).
The errors on these profiles were calculated by combining the uncertainties of each elliptical annulus with the one of the local sky estimate, as is outlined in \cite{2024Golini}.

\begin{figure}
    \centering
    \includegraphics[width=0.45\textwidth]{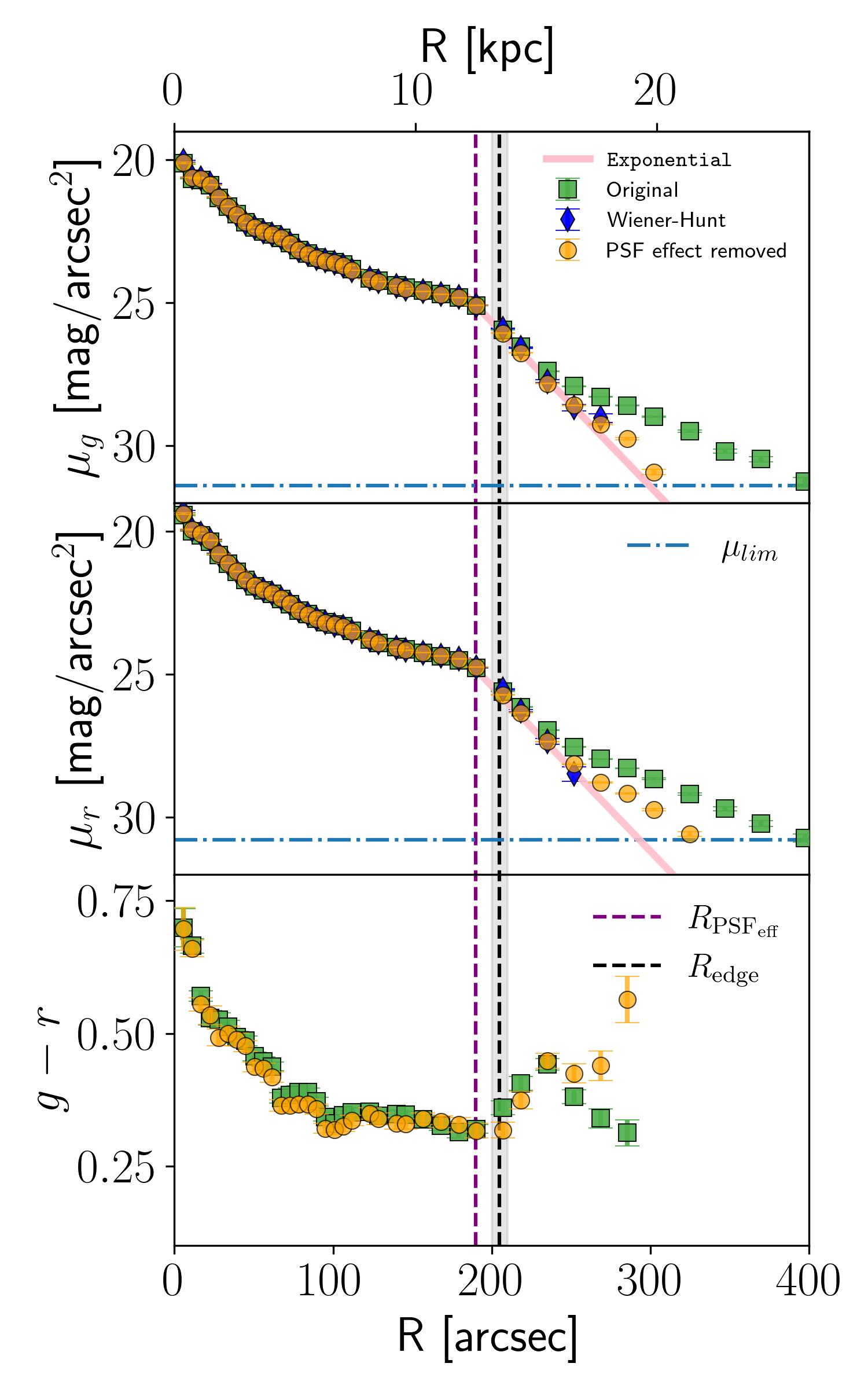}
    \caption{Surface brightness profiles of NGC 3486 in the $g$ band (upper panel) and $r$ band (central panel) using LIGHTS data. The surface brightness profiles shown correspond to the background-subtracted data (green), the Wiener-Hunt deconvolved data (blue), and the data for which the PSF effect has been removed (orange). These profiles were obtained using a fixed ellipticity and PA representative of the outer parts of the galaxy (see the main text for details). Only values above 3$\sigma$ the noise of the data are shown. The pink lines correspond to the IMFIT models we used to model the Wiener-Hunt deconvolved light distribution of the galaxy. The vertical dashed black lines highlight the location of $R_\mathrm{edge}$ with the corresponding shaded gray uncertainty, dashed vertical purple lines highlight the location of $R_\mathrm{PSF_{eff}}$ (the radius where the PSF starts to be noticeable), and horizontal blue lines represent the data surface brightness limiting depth. The $g-r$ average radial profiles prior and after deconvolution are shown in the bottom panel.}
    \label{fig:sb-lbt}
\end{figure}

In Fig. \ref{fig:sb-lbt}, we show in green the resulting $g$ and $r$ surface brightness profiles prior to deconvolution, which exhibit an inner component extending to about 40$\arcsec$, followed by a exponential decline. At around 200$\arcsec$, a slope change occurs, with the profile steepening in both bands, marking the edge of the visible disk seen in Fig. \ref{fig:ngc3486rgb}. 

The Wiener-Hunt deconvolved surface brightness profiles in the $g$ and $r$ bands are also shown in Fig. \ref{fig:sb-lbt}.
Compared to the original (background-subtracted) profiles, the deconvolved profiles exhibit a steeper outer slope beyond R = 190$\arcsec$. We visually define this transition point as $R_{\mathrm{PSF_{eff}}}$. 
This effect is due to the redistribution of light from the galaxy's outskirts to the inner regions, compensating for the PSF-induced broadening in the original data. For instance, post-deconvolution, the inner region (R < 20$\arcsec$) appears 0.1 magnitude brighter, while outer regions (R > $\Rpsf{}$) are up to 1 magnitude fainter. The edge location is preserved.

The Wiener-Hunt filter is a widely used deconvolution technique, but like other deconvolution methods such as Richardson-Lucy \citep{1972Richardson,1974Lucy} or ones based on spatial regularizations such as the assumption of sparsity in a wavelet domain \citep{starck1994image,2006Starck}, it introduces image smoothness, correlates pixel values, and alters background statistics, especially in regions of LSB \citep{murli1999wiener,Orieux,ramos2020analysis, yoo2014image}.

Despite this drawback, the Wiener-Hunt filter remains robust to saturation effects, PSF model inaccuracies, and varying background noise levels when averaging surface brightness distributions. These tests are illustrated in Appendix \ref{app:wavelettests} along with a comparison to wavelet-regularized deconvolution.
However, due to the effects of the deconvolution methods on the LSB structures, the treatment of the outermost region of the galaxy requires a more sophisticated procedure, which we explain below.

\subsubsection{Creation of a model of the galaxy not affected by PSF}
\label{subsub:modelinglight}

To remove the PSF effect from the NGC 3486 data, while preserving the noise structure of the LIGHTS image, we proceeded as follows.
First, we modeled the averaged surface brightness profile of NGC 3486 in both bands after Wiener-Hunt deconvolution (blue profiles in Fig. \ref{fig:sb-lbt}) using a custom implementation of the IMFIT \texttt{Exponential} function in Python.

We modeled the Wiener-Hunt deconvolved profile
in the range between $R_\mathrm{PSF_{eff}}$ (vertical dashed purple lines in Fig. \ref{fig:sb-lbt}), which is the point at which the PSF begins to significantly affect the outermost regions of the light profiles in the $g$ and $r$ filters (i.e., $>$ 25 mag/arcsec$^2$), to the radius where the uncertainty reaches 0.1 mag. This results in an outer exponential scale length ($h_2$) of 18.8$\arcsec$ in both filters.
Second, we generated a 2D image of this \texttt{Exponential} model using \texttt{makeimage}.
When constructing the 2D outer disk component, we adopted the PA and inclination parameters obtained in Appendix \ref{stellar-geometry}.
We did not include a bulge or stellar halo component to avoid imposing a priori assumptions. Stellar halos often exhibit asymmetries, spatial offsets, or orientations misaligned with the disk, making them difficult to model reliably.
Our final model of the galaxy, not affected by the PSF, consists of two parts: an inner region (R < $R_\mathrm{PSF_{eff}}$) represented by the Wiener-Hunt deconvolved image, which enhances spiral structures that would otherwise be difficult to model, and an
outer region (R > $R_\mathrm{PSF_{eff}}$) that transitions to an exponential declining profile (shown in pink in Fig. \ref{fig:sb-lbt}).

\subsubsection{Subtraction of the model and residuals addition: PSF deconvolved data}
\label{subsub:psf-dec-data}

\begin{figure*}[h!]
    \centering
    \includegraphics[width=\linewidth]{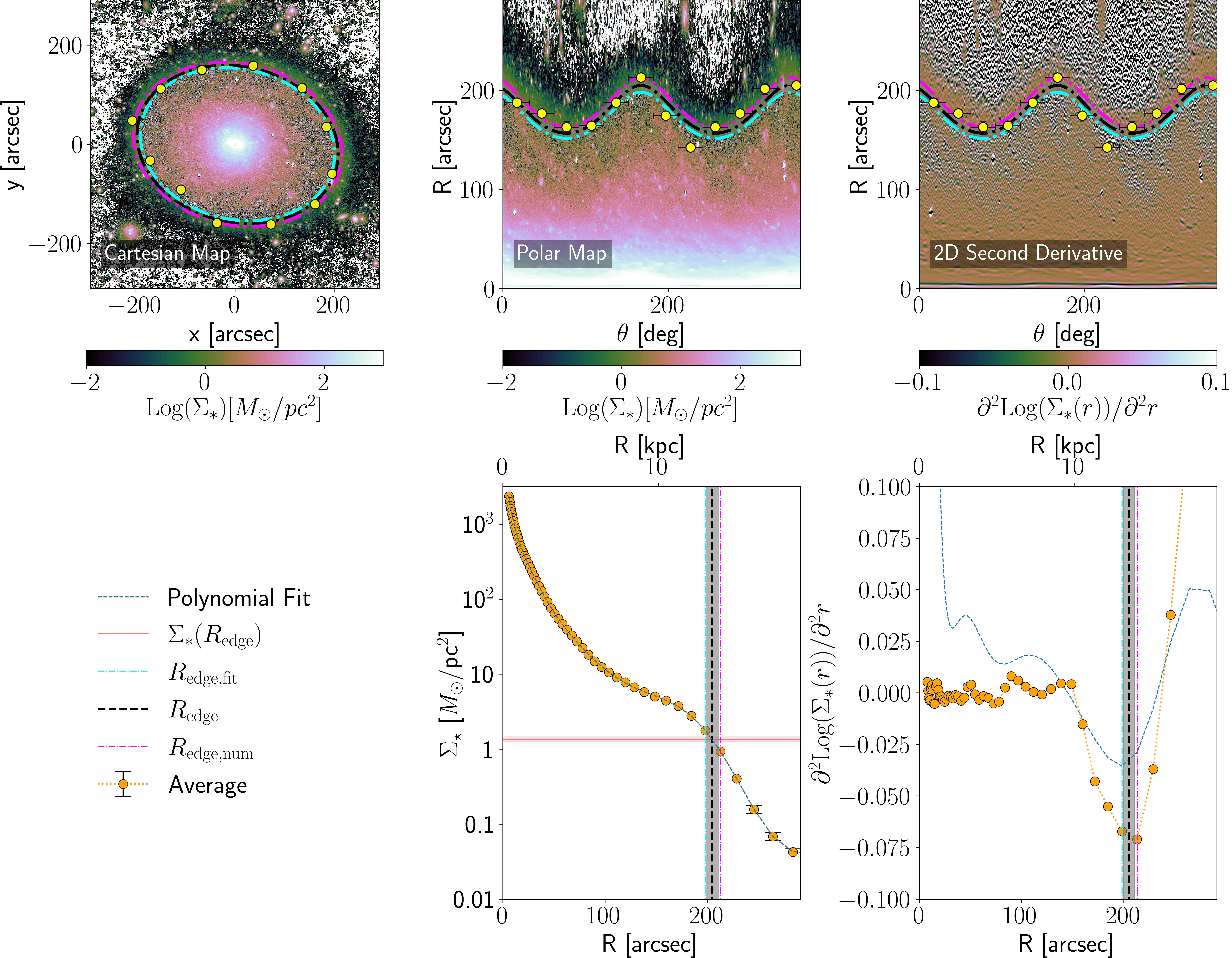}
    \caption{Edge detection in NGC 3486.
    Upper row: Stellar mass surface density map in Cartesian and polar coordinates (left and central panels) and polar map of its second derivative (right).
    Bottom row: Radial profile of averaged stellar mass surface density as a function of radius (central panel) along with its second derivative (right panel). A polynomial fit to the data and its second derivative are also shown (blue). In all panels shaded regions indicate measurement uncertainties. The outer boundary, $R_\mathrm{edge}$ is indicated in all panels by a dashed black line. $\Sigma_{*}(R_\mathrm{edge})$ is shown as a horizontal red line in the bottom central panel. Yellow dots in the upper panels mark $R_{\textnormal{edge}, \theta}$.}
    \label{edgedetectionlbt}
\end{figure*}

After generating the two deconvolved models (one for $g$ and one for the $r$ band) of NGC 3486, we employed the \texttt{astconvolve} tool in Gnuastro to convolve these models with the LIGHTS' PSF. To isolate structural components beyond the PSF effect, we subtracted these convolved images from the original galaxy data using the \texttt{astarithmetic} tool in Gnuastro, resulting in a “residuals” image. 
These residuals highlight the galaxy’s asymmetric faint features in the outer region of the galaxy that are not accounted for in the symmetric model.

Following this, we added the residuals image back into the model created in Sect. \ref{subsub:modelinglight}, resulting in a refined image with PSF effects effectively removed.
This stage of analysis ensures that any faint residual light detected in the galaxy’s outer regions corresponds to real, low-luminosity stellar structures rather than artifacts introduced by the PSF.

The surface brightness profiles, after PSF correction, extracted using fixed ellipses, are shown in orange in Fig. \ref{fig:sb-lbt}. In both bands, the galaxy light distribution after PSF removal shows a steeper decline, up to $\sim$ one magnitude fainter, starting at R $\sim$ $\Rpsf{}$.
We remark that, to avoid contamination, all surface brightness profiles shown in Fig. \ref{fig:sb-lbt} were computed after applying a mask (see Fig. \ref{fig:mask}) to the images.

\subsection{$g-r$ color and surface stellar mass density}
\label{stellar-mass}

With PSF effects removed, the galaxy's outer light distribution now reliably represents the intrinsic stellar light of the galaxy and we can proceed to the analysis of the $g-r$ color and stellar mass density profiles. The averaged $g-r$ color profile (shown in the lower panel of 
Fig. \ref{fig:sb-lbt}) of NGC 3486 shows a typical U-shaped profile \citep{Bakos2008, 2008azzollini} with an inner red bulge ($g -r $ $\sim$ 0.7) transitioning to a bluer disk with an average $g -r $ $\sim$ 0.3 at R = 100$\arcsec$ and then reddening to $g -r $ $\sim$ 0.45 beyond R = 230$\arcsec$.

We derived the 2D surface mass density map of NGC 3486 using the mass-to-light ratio (M/L)$_{\lambda}$ versus color relation prescribed in \cite{2015roediger}. Explicitly we used the $g$ Sloan data as reference as it is our deepest imaging dataset and the $g-r$ color. We assumed a Chabrier initial mass function (IMF; \citealt{chabrier2003}). 
The color image was created by subtracting the g and r surface brightness 2D maps and correcting for Galactic extinction using the following coefficients: $A_g$ = 0.08 and $A_r$ = 0.06 \citep{2011schlafly}.
Inclination correction was included following the prescriptions in \cite{2020Trujillo}.
The resulting Cartesian and polar maps of the stellar surface mass distribution of NGC 3486 are shown in the top left and top central panels of Fig. \ref{edgedetectionlbt}, respectively.

After constructing the stellar surface mass density map, we applied the mask generated in Sect. \ref{sky-mask} to exclude sources not belonging to the galaxy. Then, to extract the radial profile, we used fixed elliptical apertures as in Sect. \ref{psf-halodeg}. The resulting stellar mass density profile is presented in orange in the bottom central panel of Fig. \ref{edgedetectionlbt}.
The profile reveals an inner luminous component, followed by a gradual exponential decline out to R $\sim$ 200$\arcsec$. Beyond this radius, the mass density profile steepens.

\subsection{One-dimensional edge detection: $R_\textnormal{edge}$}
\label{1d}

We computed the numerical derivatives of the averaged stellar mass density profile of NGC 3486. The minimum of the second numerical derivative lies at $R_{\mathrm{edge,num}} = 213\arcsec$ $\pm$ 5$\arcsec$\footnote{The error associated with the numerical derivative is half of the width of the bin we use to extract the profile.}. 
To test the result, we fit a polynomial function to the average stellar mass density profile and we computed its analytical derivatives (its second derivative is shown as a dashed blue line in the bottom right panel of Fig. \ref{edgedetectionlbt}).
The minimum obtained from the polynomial fit occurs at $R_\mathrm{edge,fit} = 198\arcsec$.
Considering the uncertainty between these two approaches, we adopted a final $R_\mathrm{edge}$ = \Redge{}, indicated by the dashed black line. 
This radius corresponds to a mean surface stellar mass density of \Sigmaredge{} $M_{\odot}/pc^2$, a $\mu_{r}(R_\mathrm{edge})$ = \muredge{} and a $\mu_{g}(R_\mathrm{edge})$ = \mugedge{}.
We discuss the implications of these values in the following sections.

We illustrate $R_\mathrm{edge}$ in two ways for a more intuitive understanding. First, in the upper left panel of Fig. \ref{edgedetectionlbt}, $R_\mathrm{edge}$ is represented as an ellipse. Second, $R_\mathrm{edge}$ is represented as a sinusoidal wave on the polar map of the galaxy (central upper panel of Fig. \ref{edgedetectionlbt}). 

In the top right panel of Fig. \ref{edgedetectionlbt}, we show the 2D map of the second derivative, representing changes in the angular direction and highlighting the natural rise of the edge of the galaxy.
The location of the edge is also shown as a vertical dashed black line in Fig. \ref{fig:sb-lbt}. 
Visual inspection tends to place the edge at a smaller radius, especially in profiles with smoother transitions between the inner and outer exponential regions (i.e., a lower $\alpha$ parameter in Eq. \ref{general_formula}). This is a perceptual bias: smoother transitions can mislead the eye into identifying a drop earlier than it actually occurs. In contrast, our edge detection method uses the second derivative to pinpoint the inflection point (see Appendix \ref{app:theory} for a visual example of this effect).

\subsection{Bidimensional edge detection: $R_{\textnormal{edge},\theta}$}
\label{2d}

In order to characterize the edge of the galaxy in two dimensions, we considered the following.
To estimate the uncertainty in the intensity profile at each radial location ($E_{\text{tot},r}$), we computed the standard deviation of the counts within an elliptical aperture $\sigma_{\text{tot}}(r)$. Under Gaussian approximation, the uncertainty in the total intensity measurement is given by $E_{\text{tot},r} = {\sigma_{\text{tot}}(r)}/{\sqrt{N_{\text{pix, aperture}}(r)}}$.
Therefore, dividing the elliptical aperture into 60-degree wedges reduces the number of pixels per wedge by a factor of 6 and increases the uncertainty by a factor of 2.5. 

This division into 60 degree wedges still provides a sufficient S/N for reliable edge detection.
Therefore, to map the edge at different angular positions ($R_{\textnormal{edge},\theta}$), we divided the galaxy into 60-degree wedges. This means that at the semimajor axis, we defined a wedge spanning PA = -30 deg to 30 deg, then shifted by 30 degrees to define another wedge from PA = 0 deg to 60 deg, and so on. This means 12 wedges in total.

\begin{figure}
    \centering
    \includegraphics[width=\linewidth]{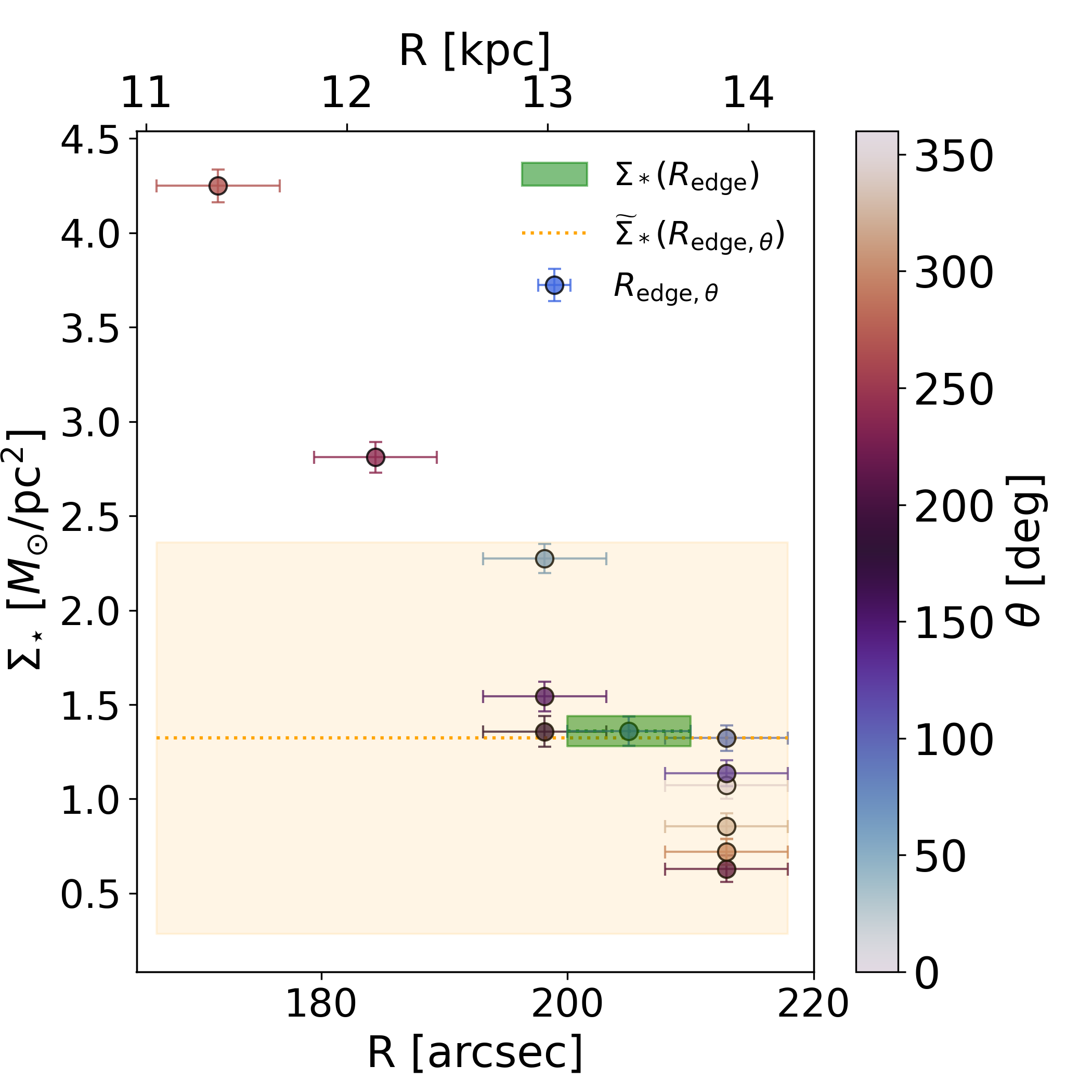}
    \caption{Distribution of stellar surface mass density at $R_{\textnormal{edge},\theta}$. The green box  corresponds to \Sigmaredge{} $M_{\odot}/pc^2$ at $R_{\textnormal{edge}}$. The orange line corresponds to the median of the $\Sigma_{\star}(R_{\textnormal{edge},\theta})$. The colored region corresponds to a 1$\sigma$ dispersion. Coincidence of locations at $R_{\textnormal{edge},\theta}$ occurs because of the limited resolution of the method. Points are color-coded by $\theta$. }
    \label{fig:stell}
\end{figure}

For each wedge, we extracted the radial profile using masked data and sampled it using the task \textit{astscript-radial-profile}.  
As in the case of the 1D approach, the final value of $R_{\textnormal{edge},\theta}$ is the mean between the minimum obtained with the numerical derivation and the analytical derivation of the polynomial fit. $R_{\textnormal{edge},\theta}$ are represented as yellow dots in the upper panels of Fig. \ref{edgedetectionlbt}. 
The errors associated with these estimates correspond to $\pm$ 5$\arcsec$ (the half width of the step).
Fig. \ref{fig:stell} shows the distribution of the $R_{\textnormal{edge}, \theta}$ and corresponding $\Sigma_{\star}$($R_{\textnormal{edge}, \theta}$) compared to the value obtained with the 1D analysis.
The 1D and 2D methods yield consistent edge values, with angular positions of $R_{\textnormal{edge},\theta}$ clustering around $R_\mathrm{edge}$ (few kiloparsec), capturing small asymmetries in the disk’s structure.
We calculated the average absolute deviation of all $R_{\textnormal{edge}, \theta}$ from $R_\mathrm{edge}$, and express this value as a percentage of $R_\mathrm{edge}$. This gives a mean asymmetry of $\sim 5 \%$ for NGC 3486.
The stellar mass densities at the edge $\Sigma_{\star}(R_{\textnormal{edge},\theta})$ cluster around the mean edge density $\Sigma_{\star}(R_{\textnormal{edge}})$, ranging between 0.75 and 4.5 $M_{\odot}/pc^{2}$.
In Fig. \ref{fig:stell}, $R_{\textnormal{edge},\theta}$ values are color-coded by their corresponding angular coordinate, $\theta$, revealing that neighboring angular sectors exhibit similar $R_{\textnormal{edge},\theta}$, suggesting local coherence in edge structure.

\section{Discussion}
\label{sec:discussion}

Our results indicate that it is possible to accurately detect the edge of galaxies through an unbiased mathematical definition, using the stellar surface mass density distribution.
NGC 3486 shows an almost symmetric disk (asymmetry < 5$\%$) with a sharp decline beyond $R_\mathrm{edge}$ = \Redge{} equal to 13.8 $\pm$ 0.6 kpc at a distance of 13.6 Mpc.
This corresponds to $\mu_{r}(R_\mathrm{edge})$ = \muredge{}, $\mu_{g}(R_\mathrm{edge})$ = \mugedge{}, a mean surface stellar mass density of \Sigmaredge{} $M_{\odot}/pc^2$, and $\Sigma_{\star}(R_{\textnormal{edge},\theta})$ values between 0.75 to 4.5 $M_{\odot}/{pc^2}$.

We note that, assuming the IMF does not vary with radius, switching to a different IMF (e.g., Salpeter) would uniformly scale the stellar mass density profile up or down. While this would affect the $\Sigma_{\star}(R_\mathrm{edge})$ estimate, it would not alter dramatically the overall shape of the profile. Consequently, the absolute values of $\Sigma_{\star}(R_{\rm edge})$ and $\Sigma_{\star}(R_{\textnormal{edge},\theta})$ reported here are IMF-dependent and would vary accordingly.

The results of this work therefore align with the ones of \citet{2022chamba-tracinglimits} for galaxies of a similar mass and redshift ($M_{*} \sim 10^{10}$ $M_{\odot}$).
They report $R_\mathrm{edge}$ values ranging from 6 to 20 kpc ( $\sim $16 kpc on average) and a stellar mass density at the edge for S0/a-Irr galaxies between 0.2 and 4 $M_{\odot}/pc^2$, with an average of $\sim $1.15 $M_{\odot}/pc^2$ (see their Fig. 5). 
The galaxies in \citet{2022chamba-tracinglimits} that have both a similar stellar mass and an average $g-r$ color to that of NGC 3486 ($\sim$ 0.3) have average edge radii greater than 10 kpc and tend to be larger than their redder counterparts.

\subsection{The formation of a sharp truncation in NGC 3486}

\begin{figure}
    \centering
    \includegraphics[width=\linewidth]{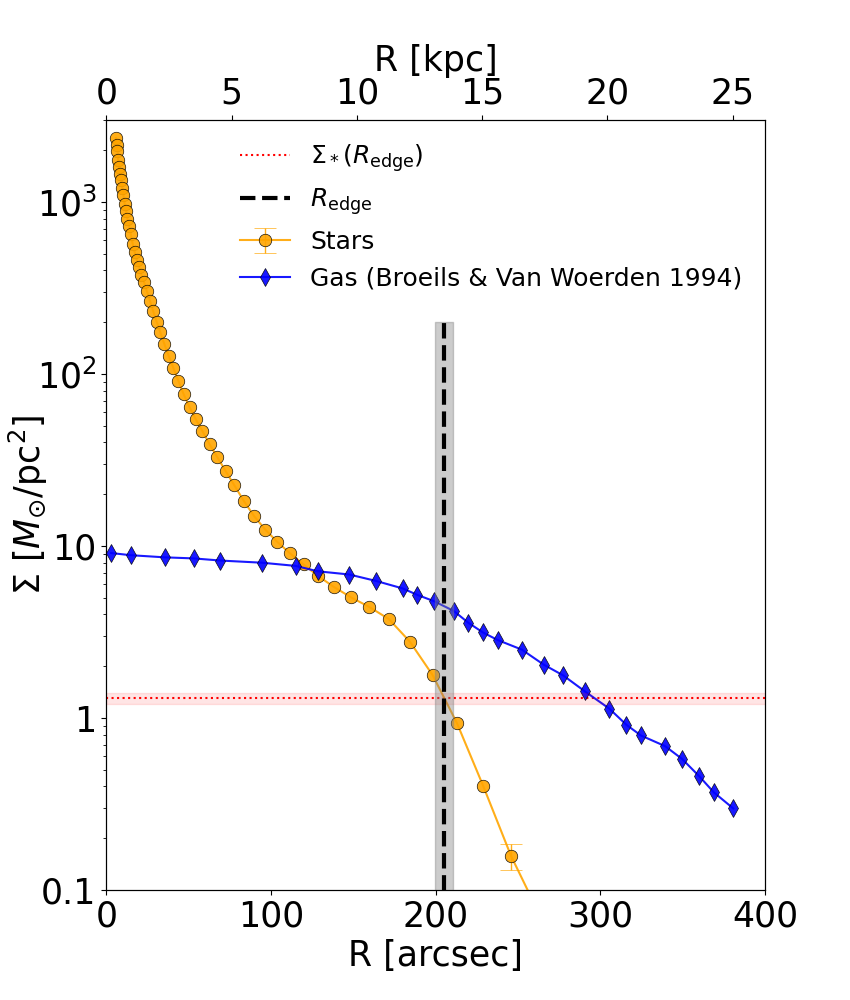}
    \caption{Surface stellar (orange; this work) and gas (blue; \citealt{broeils1994search}) mass density profiles of NGC 3486. The stellar mass density profile shows an edge feature at \Redge{} (gray band), which coincides with the radius at which the gas profile also decays more rapidly. We assume a distance of 13.6 Mpc \citep{kourkchi2020cosmicflows}. $\Sigma_{\star}(R_{\rm edge})$ is marked with a horizontal red line.}
    \label{fig:RADIOMAP}
\end{figure}

NGC 3486 shows a broken exponential stellar surface mass density profile, characterized by a change in slope that separates the decline of the inner region from the steeper outer regions.  This transition appears sharp, with a ratio of $h_{2}/h_{1} \sim$ 5 after removal of the effect of the PSF. This sharp transition is similar to that observed in the edge-on galaxies NGC 5907 and NGC 4565 \citep{Lombilla_2018}.
The near-ultraviolet (NUV) and far-ultraviolet (FUV) profiles (see Fig. \ref{fig:all-sbl}) also show a sharp drop, probably indicating negligible star formation beyond the edge. A similar break is also suggested in the infrared profile from \cite{2016Richardsh1}, where a change in slope is observed at $\sim$ 200$\arcsec$ (see their Fig. A.1).

The presence of this clear edge suggests an ongoing star formation process. Over time, this sharp boundary is expected to fade as stellar migration redistributes stars from the inner regions inside $R_\mathrm{edge}$ into the outer disk \citep{2014Sellwood, 2017Debattista}, gradually erasing signatures of past star formation.
Further evidence for this star formation link can be found in the $g-r$ color profile. Within $R_\mathrm{edge}$ the galaxy shows active star formation, while beyond this radius the $g-r$ color becomes redder, possibly due to stellar migration. This is consistent with the results of \cite{Lombilla_2018}, who interpret the gradual reddening beyond the edge as passive aging of the stellar population.

Under the hypothesis that $R_\mathrm{edge}$ marks the limit of in situ star formation, the formation and evolution of the edge is intimately linked to the abundance of HI gas. The absence of star formation in the outer disk is likely tied to the availability of cold gas and feedback effects. Observations of HI and molecular gas distributions in galaxies with similar profiles show that gas densities drop below the critical threshold for star formation beyond the edge. Using Very Large Array (VLA) data, \cite{2016Richardsh1} found that NGC 3486 has an asymmetric HI gas distribution that extends from the northwest and southeast sides of the HI disk. They also infer a mean inclination from the HI data at 200$\arcsec$ of $\sim 40$ deg that is consistent with the AR of 0.765 found in this work.
In Fig. \ref{fig:RADIOMAP-APP} (see Appendix \ref{App:hi-map}) we have tentatively superimposed the HI contours from \cite{2016Richardsh1} on the LIGHTS data.
The asymmetric HI gas distribution is further analyzed by \cite{2021smith}, who reports two HI arcs extending from the southeastern and northwestern parts of NGC 3486 (see their Fig. A.9) and rotating with the rest of the galaxy.
We hypothesize that the observed HI asymmetry could be due to an interaction with a satellite galaxy (PGC 033184; see Fig. \ref{fig:ngc3486rgb}) located to the north of NGC 3486 at z = 0.008. 
The gas content of NGC 3486 was also studied by \cite{broeils1994search} using short 21-cm line observations from the Westerbork Synthesis Radio Telescope.  
Fig. \ref{fig:RADIOMAP} shows a comparison between their derived surface gas mass density profile (blue) and the surface stellar mass density of this work (orange). We have adapted their profile to the distance assumed in this paper.  
At \( R_\mathrm{edge} \), the surface HI gas mass density reaches $\sim$ \( 5 \, M_{\odot}/pc^2 \), consistent with observations of nearby spiral galaxies \citep{martin2001star,leroy2008star}.  
In NGC 3486, the star formation efficiency at \( R_\mathrm{edge} \) is around 25$\%$, as the surface gas mass density at this radius is $\sim$ 3.8 times higher than the stellar mass density.
This result is therefore consistent with theoretical models suggesting that the star formation threshold in galaxies is regulated by thermal instabilities arising from the transition between warm and cold gas phases \citep{2004Schaye}.  

\subsection{Method limitations and future directions}
\label{subsec:limitations}

We evaluated the performance and implications of our new edge detection method, which mitigates the effects of scattered light and helps to identify the edge location (and therefore the characterization of the size using this feature as a proxy) in galaxies. This method is crucial in deep imaging, providing a robust approach to identifying galaxy edges based on second derivatives. This technique also provides the first wedge-based characterization of the border of galaxies.
The code is modular and automated, including profile extraction, derivative calculation, and edge detection in both 1D and 2D. 
We acknowledge that  masking LSB features, deciding on the sampling of the surface brightness and mass profiles, or adapting wedge widths still involves manual decisions and may carry subjective elements. Computing the second derivative can introduce numerical errors, which may stem from point spacing or limited floating-point precision. Our approach minimizes smoothing to preserve data authenticity, but this remains a balancing challenge.
While computationally efficient overall, the most resource-intensive step remains the PSF correction of the data. 
Future work (Golini et al., in prep.) will explore the method’s reproducibility across datasets with varying depths, noise levels, resolution (i.e., different sampling of the mass profile), and PSF models, and expand its application to a broader range of galaxy properties (e.g., inclination, size, mass).

\section{Conclusions}
\label{sec:conclusions}

We have introduced a robust method of measuring the edges of galaxies by leveraging the second derivative of the stellar surface mass density profile. This approach serves as a complementary quantitative alternative to traditional visual-based methods \citep{Lombilla_2018, 2020Trujillo, 2022Diazgarcia, 2022chamba-tracinglimits, Buitragotrujillo2024} of identifying galaxy boundaries. Our technique enables both 1D and 2D analysis, allowing for a comprehensive characterization of galaxy outskirts, asymmetries, and structural evolution. By applying our method to deep imaging observations of NGC 3486 from the LIGHTS \citep{LIGHTSs, 2024Zaritskydennislights} survey ($\mu_{r,lim}$ = \rsblimLBT{} mag/arcsec$^2$ and $\mu_{g,lim}$ = \gsblimLBT{} mag/arcsec$^2$; 3$\sigma$ in 10$\arcsec$ $\times$ 10$\arcsec$ boxes), we identified an outer edge at \Redge{}, in strong agreement with the visual boundary. Our wedge-based analysis quantified the galaxy’s asymmetry at 5$\%$, confirming the consistency between the isophotal ellipse parameters and the edge contours at different angular directions. The sharp transition at the edge occurs at a stellar surface mass density of $\sim 1 M_{\odot}$/pc$^2$, reinforcing the possible link between galaxy edges and the threshold of in situ star formation \citep{1972Quirk, 1989Kennicutt, 2004Schaye, 2008rokroskar}.

The computational efficiency and adaptability of this method make it well suited for integration into survey pipelines, ensuring consistent edge detection across different datasets, such as LSST \citep{2019lsst}, Euclid \citep{scaramella2022euclid}, Roman \citep{bailey2023nancy}, and ARRAKIHS \citep{guzman2024arrakihs}.
In summary, this method offers a promising advance in galaxy edge detection, while aligning well with the needs of deep imaging surveys (such as background and scattered light removal). Continued development will enable its application across varied datasets, paving the way for an improved understanding of galaxy structures and the processes that shape them.

\begin{acknowledgements}
We thank the referee for the constructive comments that helped improve the quality of the manuscript.
GG acknowledges support from the PID2022-140869NB-I00 grant from the Spanish Ministry of Science and Innovation. IT acknowledges support from the ACIISI, Consejer\'{i}a de Econom\'{i}a, Conocimiento y Empleo del Gobierno de Canarias and the European Regional Development Fund (ERDF) under a grant with reference PROID2021010044 and from the State Research Agency (AEI-MCINN) of the Spanish Ministry of Science and Innovation under the grant PID2022-140869NB-I00 and IAC project P/302302, financed by the Ministry of Science and Innovation, through the State Budget and by the Canary Islands Department of Economy, Knowledge, and Employment, through the Regional Budget of the Autonomous Community. This research also acknowledge support from the European Union through: "UNDARK" and "Excellence in Galaxies - Twinning the IAC" of the EU Horizon Europe Widening Actions  programmes (project numbers 101159929 and 101158446) and (MSCA EDUCADO, GA 101119830). Views and opinions expressed are however those of the author(s) only and do not necessarily reflect those of the European Union or European Research Executive Agency (REA). Neither the European Union nor the granting authority can be held responsible for them.
RIS acknowledges financial support from the Spanish Ministry of Science and Innovation through the project PID2022-138896NA-C54.
MM acknowledges support from grant RYC2022-036949-I financed by the MICIU/AEI/10.13039/501100011033 and by ESF+, and program Unidad de Excelencia Mar\'{i}a de Maeztu CEX2020-001058-M. SR acknowledges support from the GEELSBE2 project with reference PID2023-150393NB-I00 funded by MCIU/AEI/10.13039/501100011033 and the FSE+ and also the Consolidación Investigadora IGADLE project with reference CNS2024-154572. SR acknowledges financial support of the Department of Education, Junta de Castilla y Le\'on and FEDER Funds (Reference: CLU-2023-1-05).
AAR acknowledges funding from the Agencia Estatal de Investigación del Ministerio de Ciencia, Innovación y Universidades (MCIU/AEI) under grant ``Polarimetric Inference of Magnetic Fields'' and the European Regional Development Fund (ERDF) with reference PID2022-136563NB-I00/10.13039/501100011033. J.R. acknowledges financial support from the Spanish Ministry of Science and Innovation through the project PID2022-138896NB-C55

\textit{Facilities}: 
LBC-Red, LBC-Blue

\textit{Software}:
Astropy             \citep{astropy1},
Astrometry.net      \citep{Lang_2010}
Gnuastro            \citep{gnuastro} 
IMFIT               \citep{Erwin2015}
Matplotlib          \citep{matplotlib} 
NumPy               \citep{numpy}
SciPy               \citep{scipy1, scipy2},
source-extractor    \citep{Sextractor}
SCAMP               \citep{2006scamp}
SWarp               \citep{swarp2010}

\end{acknowledgements}

\bibliographystyle{aa}
\bibliography{aa55288-25.bib}

\begin{appendix}
\section{Profiles of NGC 3486 at different wavelengths}

\begin{figure}
    \centering
    \includegraphics[width=\linewidth]{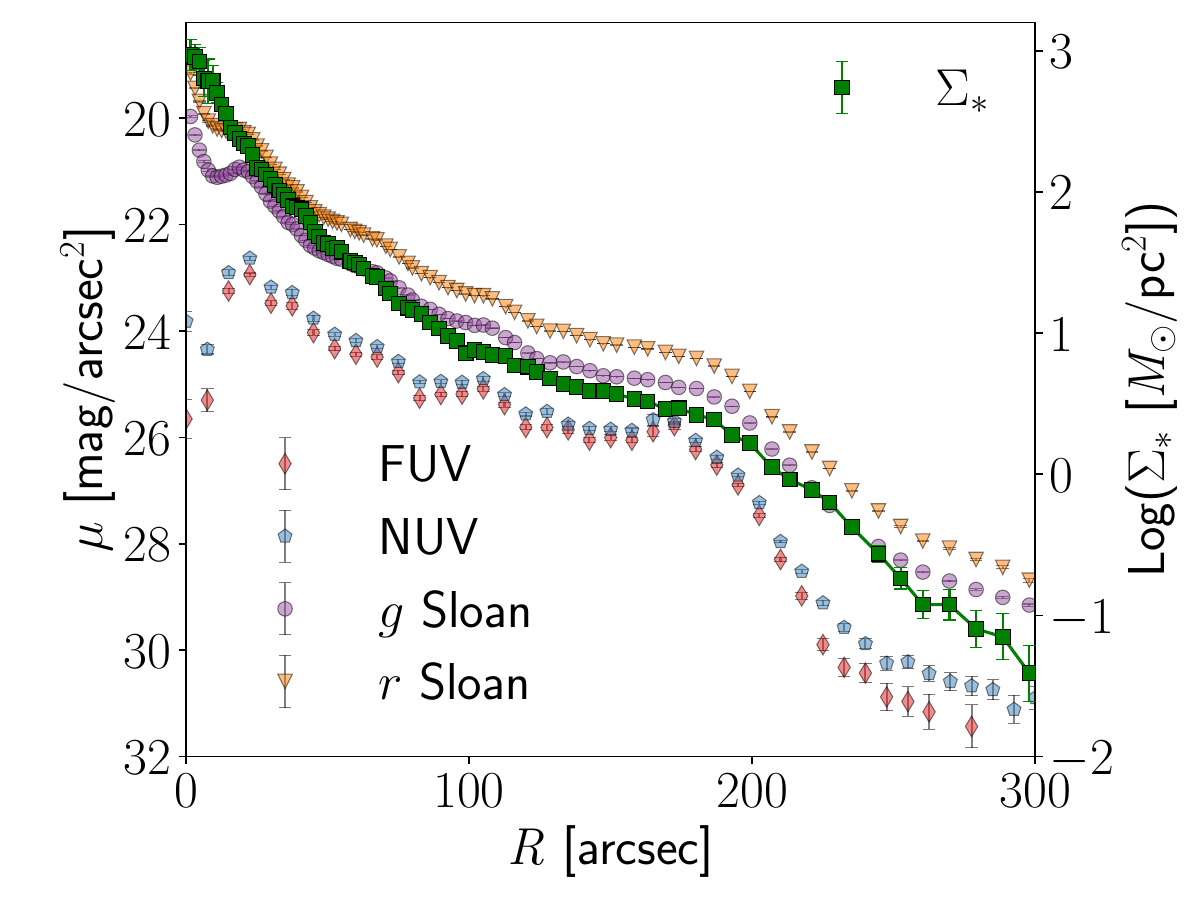}
    \caption{Light profiles of NGC 3486 using different filters. Optical and stellar mass density profiles are derived using LIGHTS data in this work. NUV and FUV profiles are taken from \cite{cejudo2025uv}.}
    \label{fig:all-sbl}
\end{figure}

We show in Fig. \ref{fig:all-sbl} the NUV and FUV average profiles of NGC 3486 derived in \cite{cejudo2025uv}, and the averaged profiles using LIGHTS data in the $g$ and $r$ bands. In the same plot, we report the stellar mass density profile ($\Sigma_{*}$) of the galaxy obtained in this work (see main text for details). $\Sigma_{*}$, on average, is smoother than the lights profiles.

\section{Color image of LIGHTS data}
\label{app:rgb}

\begin{figure*}[h!]
    \centering
    \includegraphics[width=\linewidth]{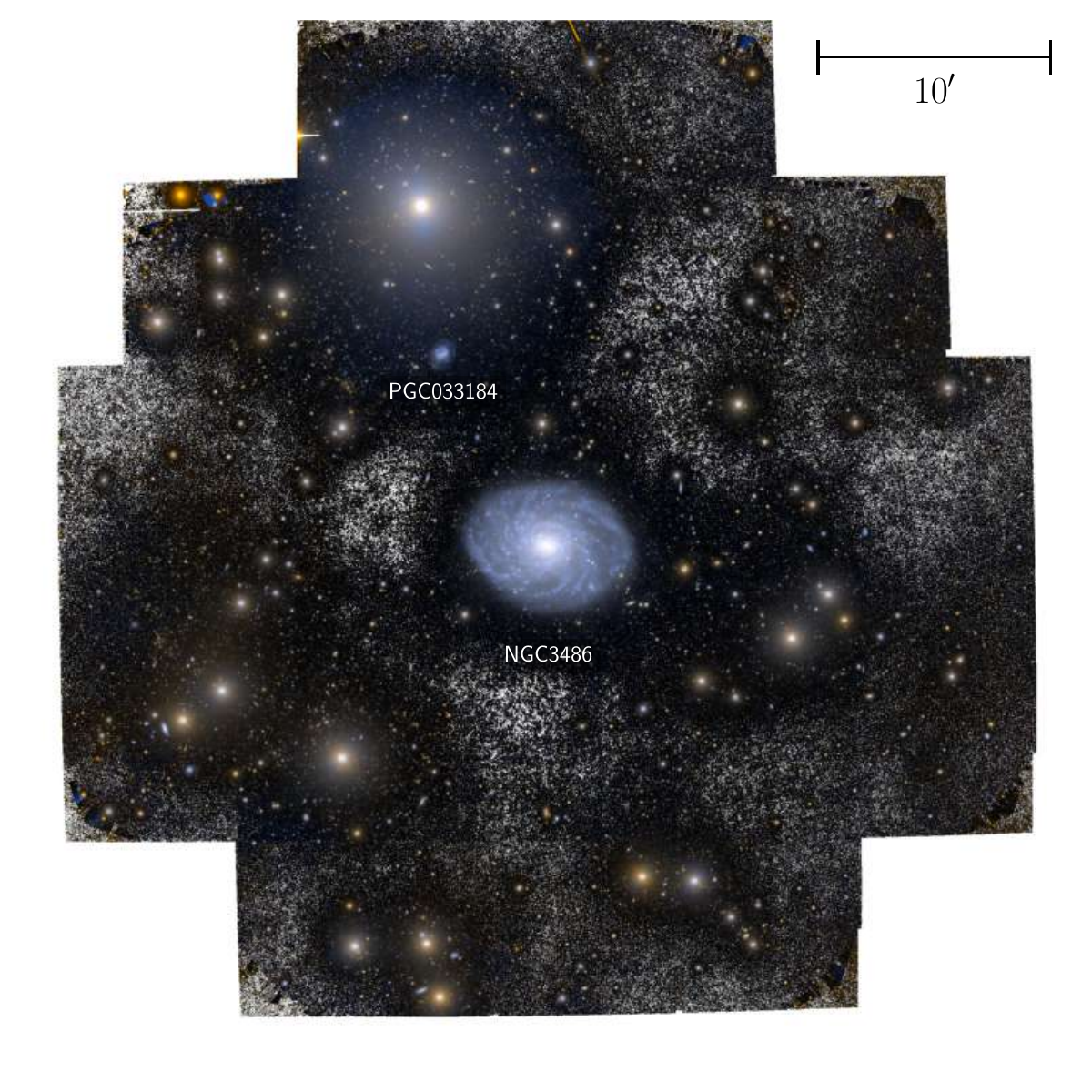}
    \caption{Color composite image of NGC 3486 and its surroundings, constructed from Sloan $g$ and $r$ band data using \textit{astscript-color-faint-gray}. Regions with low S/N appear white. The black and white background is from the $g$ band. North is up and east is to the left. At a distance of 13.6 Mpc, 10$^{\prime}$ correspond to 39.4 kpc.}
    \label{fig:ngc3486rgb}
\end{figure*}

Fig. \ref{fig:ngc3486rgb} shows the LIGHTS data of NGC 3486. The color image with gray-scaled background was constructed using \textit{astscript-color-faint-gray} \citep{scriptcolorimagesRaul}, by combining the $g$ and $r$ filters. The black and white background is from the $g$ band.\\

\section{Mask}
\label{app:maskimage}

\begin{figure}[!h]
    \centering
    \includegraphics[width=\linewidth]{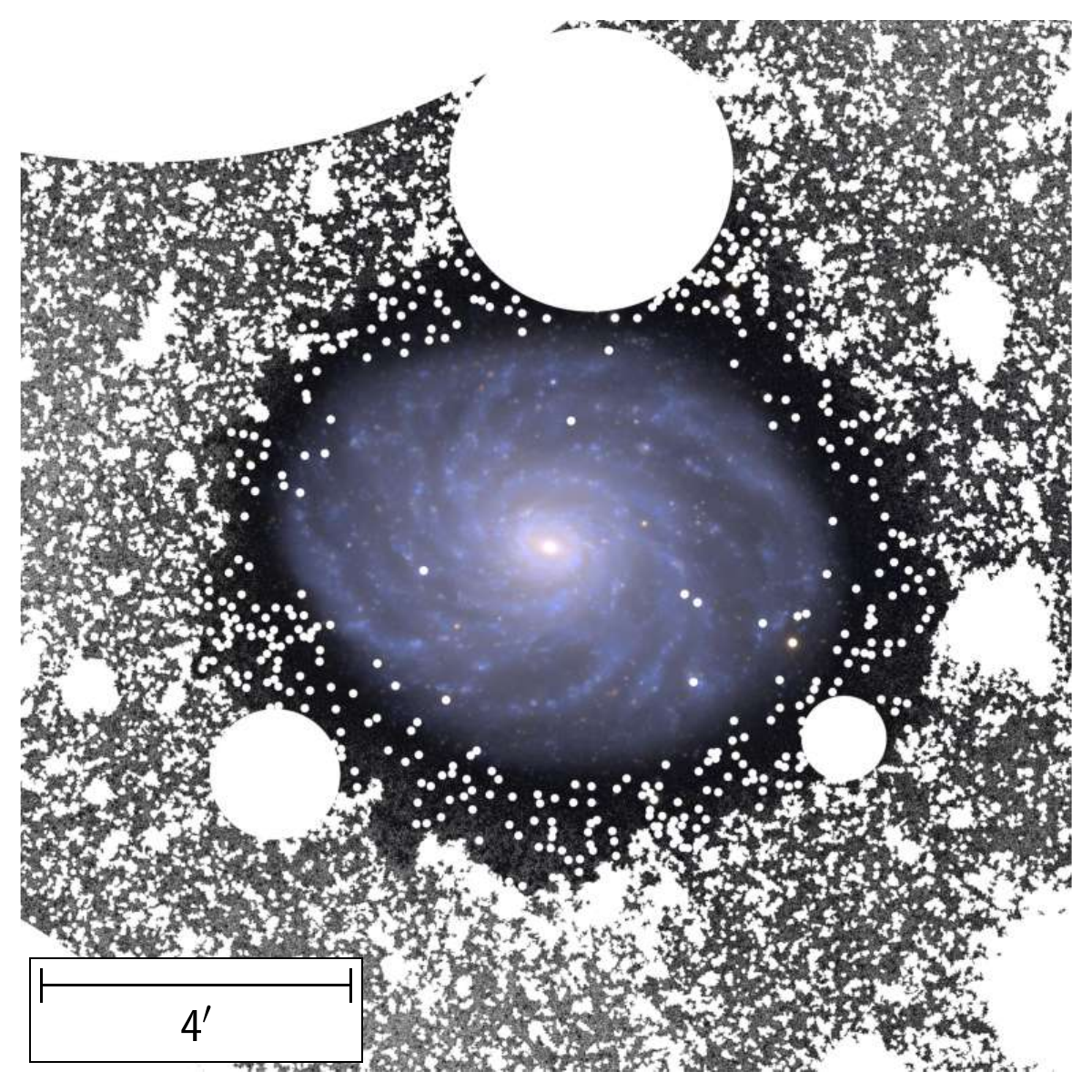}
    \caption{Mask of NGC 3486 used in this work. The mask is created by combining \texttt{Noisechisel}, \texttt{Segment}, unsharp masking and manual masking based on $g$ and $r$ band data of the LIGHTS Survey.}
    \label{fig:mask}
\end{figure}

Fig. \ref{fig:mask} shows the mask used in this work for the LIGHTS data of NGC 3486 created using a combination of $g$ and $r$ masks.

\section{Structural parameters of Model1 and Model2}
\label{app:modelsdegeneracy}

The IMFIT parameters of the simulated Model1 and Model2 share similarities but differ in key aspects. To create both models we used a \texttt{BrokenExponential} profile with a PA of -20 degrees (clockwise from the X axis), an inclination of 39.6 degrees and an inner scale length $h_1$ of 75 pixels. The location of the transition between inner and outer exponential declines was set to 200 pixels, and $\alpha$ is equal to 10 in both models.
Differences arise in the outer scale length $h_2$, where in Model1 it is set to 30 pixels, while in Model2 to 25 pixels. Additionally, in the halo \texttt{Exponential} component, Model2 has central intensity value of 0.0004 counts/pix and scale length of 270 pixels.

\section{Characterising the position angle and axis ratio of NGC3486}
\label{stellar-geometry}

To extract the surface brightness profile of NGC 3486, we first used the \texttt{ellipse} routine in \texttt{photutils} \citep{Jedrzejewski1987, Bradley2019} to fit elliptical isophotes to the combined $g$ and $r$ band data. This method provides a representative PA and ellipticity of the outer galaxy regions. In the initial run of \texttt{ellipse}, all parameters varied freely. In the subsequent run, we fixed the center to the median coordinates derived in the first pass (R.A. (J2000) = 165.0994 deg, Dec. (J2000) = 28.9751 deg)\footnote{The uncertainty associated with the selection of the center of the galaxy is $\pm$ 0.3$\arcsec$ in R.A. and Dec \citep{Jedrzejewski1987}.}. 
The fitting began at 2$\arcmin$ with linearly spaced elliptical annuli. We found an average AR of $0.765 \pm 0.020$ and PA of $13.0 \pm 0.7^\circ$ (clockwise from the X axis) between radii of 170$\arcsec$ and 220$\arcsec$. These values align with prior measurements, such as AR =  0.792 and PA = $8^\circ$ (clockwise from the X axis; \citealt{2021smith}) and PA = $5^\circ$ (clockwise from the X axis) with an inclination of 36.8 from gas kinematics \citep{2016Richardsh1}.

\section{Tests on Wiener deconvolution}
\label{app:wavelettests}

\textbf{PSF, saturation and noise.} To evaluate the accuracy of the Wiener-Hunt deconvolution method, we applied it to the simulated disk Model1. By comparing the deconvolved images to their original (not convolved) counterparts, we assessed how well the method retrieves the intrinsic averaged surface brightness profiles and we visually compared the background variations. The profiles were extracted using fixed ellipses for each model. Our findings are discussed below.

\begin{figure*}
    \centering
    \includegraphics[width=\textwidth]{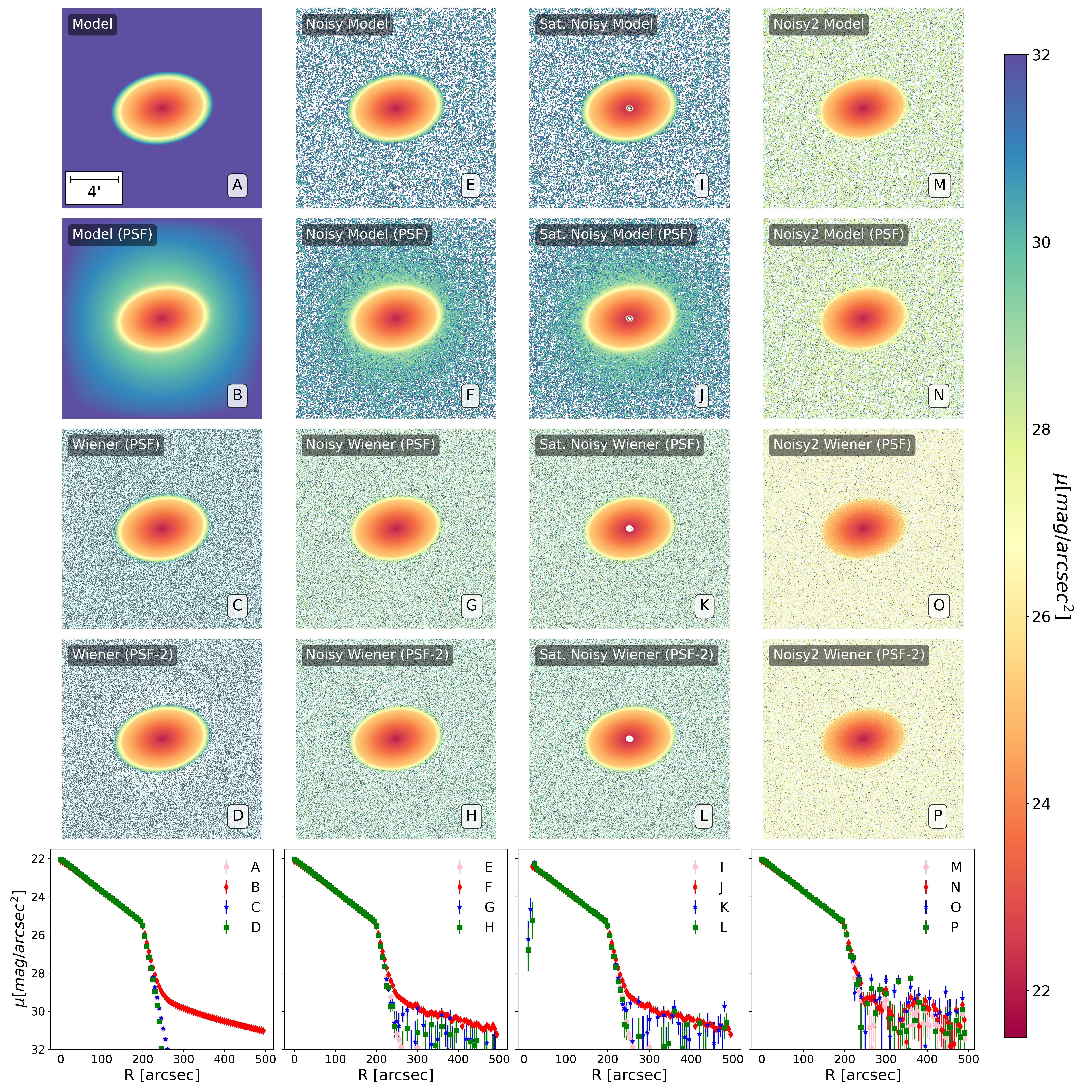}
    \caption{Effectiveness of Wiener-Hunt deconvolution on a simulated disk in case of inaccurate PSF modeling, saturation and background noise (see main text for details).} 
    \label{test-wiener}
\end{figure*}

Fig. \ref{test-wiener} presents a comparative analysis of disk models, illustrating the effects of the PSF and Wiener-Hunt deconvolution under different conditions. The first column represents the base model, while subsequent columns introduce noise and saturation effects. Each row shows a different processing stage.
The first row displays the original model (assuming infinite S/N), the same model with added Gaussian ($\sigma$ = 0.003) noise using the task \texttt{mknoise-sigma-from-mean} function in Gnuastro (Noisy model), the Noisy model with inner 20$\arcsec$ replaced with Not a Number (NaNs) simulating saturation (Sat. Noisy model), and the original model adding higher Gaussian ($\sigma$ = 0.03) noise (Noisy2 model).
The Noisy model simulates an image with a limiting surface brightness of $\mu_{lim} \sim$ 29.1 mag/arcsec$^2$ (3$\sigma$; 10$\arcsec$ $\times$ 10$\arcsec$), while the Noisy2 model corresponds to $\mu_{lim} \sim$ 26.6 mag/arcsec$^2$ (3$\sigma$; 10$\arcsec$ $\times$ 10$\arcsec$). These values are representative of typical surveys such as Stripe 82 and SDSS.

In the second row, the same models were convolved with the LIGHTS' PSF \citep{2024sedighi} using the \texttt{--psf} option in IMFIT.
The PSF declines as a power law ($r^{ \beta}$) with an exponent $\beta$ = -2.1.
On the third and forth row, the panels display Wiener-Hunt deconvolution applied with the correct PSF and with a not optimal PSF model (PSF-2) with $\beta = {-2.2}$.
In the fifth row, we show the averaged surface brightness profiles corresponding to each column, comparing the different processing steps.
The colorbar indicates the surface brightness scale. Fig. \ref{test-wiener} demonstrates how, after Wiener-Hunt deconvolution, the images introduce correlated noise in the outskirts while preserving the edge location and, how Wiener-Hunt deconvolution, even in case of noise and saturation, can partially recover the original structure, down to the limiting depth of the image.\\

\textbf{Comparison with wavelet-regularized deconvolution.} Wiener-Hunt deconvolution is not the only method we can use to address the PSF-Stellar halo degeneracy. We compared Wiener-Hunt deconvolution with wavelet-regularized deconvolution techniques. Before delving into this comparison, we introduce the concept of wavelets as follows.  
The quality of the deconvolved images can be improved by performing a regularized deconvolution. If the noise term in the image formation assumption of Eq. \ref{eq:image_formation} is Gaussian with zero mean and diagonal covariance matrix, the deconvolution can be performed in a Bayesian framework by optimizing the following metric:  

\begin{equation}
    L = \sum_{x,y} \left\| g(x, y) - f(x, y) * h(x, y) \right\|^2 + \lambda R(f),
\end{equation}  

where \( \left\| f \right\|_2 \) is the \( \ell_2 \) norm\footnote{The \( \ell_n \) norm of a vector is given by \( \parallel \mathbf{x} \parallel_n = (\sum_i |x_i|^n)^{1/n} \) if \( n > 0 \).}.  

\begin{figure}
    \centering
    \includegraphics[width=\linewidth]{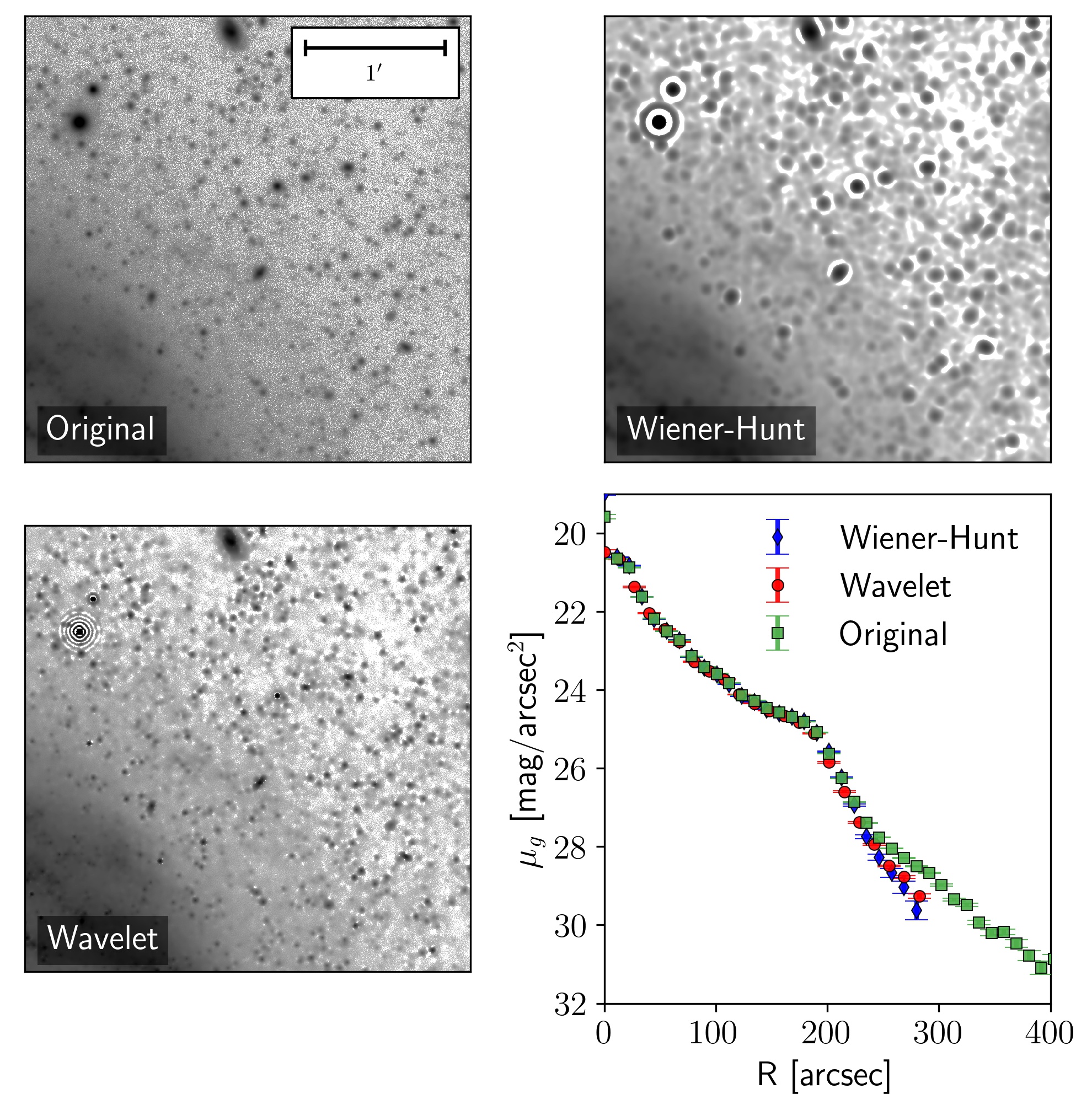}
    \caption{Effect of different deconvolution methods on the background of LIGHTS data. Gray-scaled images correspond to the $g$ band. The bottom right panel shows the averaged surface brightness profile of NGC 3486 after applying Wiener-Hunt (blue) and wavelet-regularized (red) deconvolution techniques, compared to the original profile (green).}
    \label{fig:wiener-waveletsb}
\end{figure}

This solution is known as the maximum a-posteriori (MAP) estimate. The first term represents the log-likelihood of the observations given the deconvolved image, while the second term encodes prior assumptions about the image structure through the functional \( R \). If no regularization is imposed (\( R = 0 \)), the solution reduces to the well-known maximum likelihood (ML) estimate, which is highly sensitive to noise. The prior acts as a regularization term, incorporating prior knowledge about the image to stabilize the solution. For example, Tikhonov regularization uses \( R(f) = \left\| f \right\|^2 \).  
In this application, we build on the ideas of sparsity-based regularization to deconvolve the images. In this case, the regularization term is given by the \( \ell_1 \) norm of a transformed version of the image:  
\begin{equation}
    R(f) = \left\| W(f) \right\|_1,
\end{equation}  
where \( W(f) \) represents the isotropic undecimated wavelet transform (IUWT), also known as the starlet transform \citep{starck1994image,2006Starck}. The IUWT is a redundant wavelet transform that decomposes an image into multiple frequency components, providing a sparse representation well-suited for astronomical images. 
For a more detailed explanation of the mathematical foundations behind wavelet-regularized deconvolution methods, we refer the reader to the following works: \citet{starck1994image, starck2007astronomical, starck2010sparse, carrillo2012sparsity}.
Fig. \ref{fig:wiener-waveletsb} shows a cropped region of the outskirts of NGC 3486 (3$^{\prime}$ $\times$ 3$^{\prime}$) before and after deconvolution and compares the averaged surface brightness profiles obtained with Wiener-Hunt and wavelet-regularized deconvolution methods using fixed elliptical apertures. We find close agreement up to 280$\arcsec$ corresponding to $\mu_g $ = 29 mag/arcsec$^2$.

\section{Stability of $R_\textnormal{edge}$ with varying $\alpha$}
\label{app:theory}

\begin{figure}
    \centering
    \includegraphics[width=\linewidth]{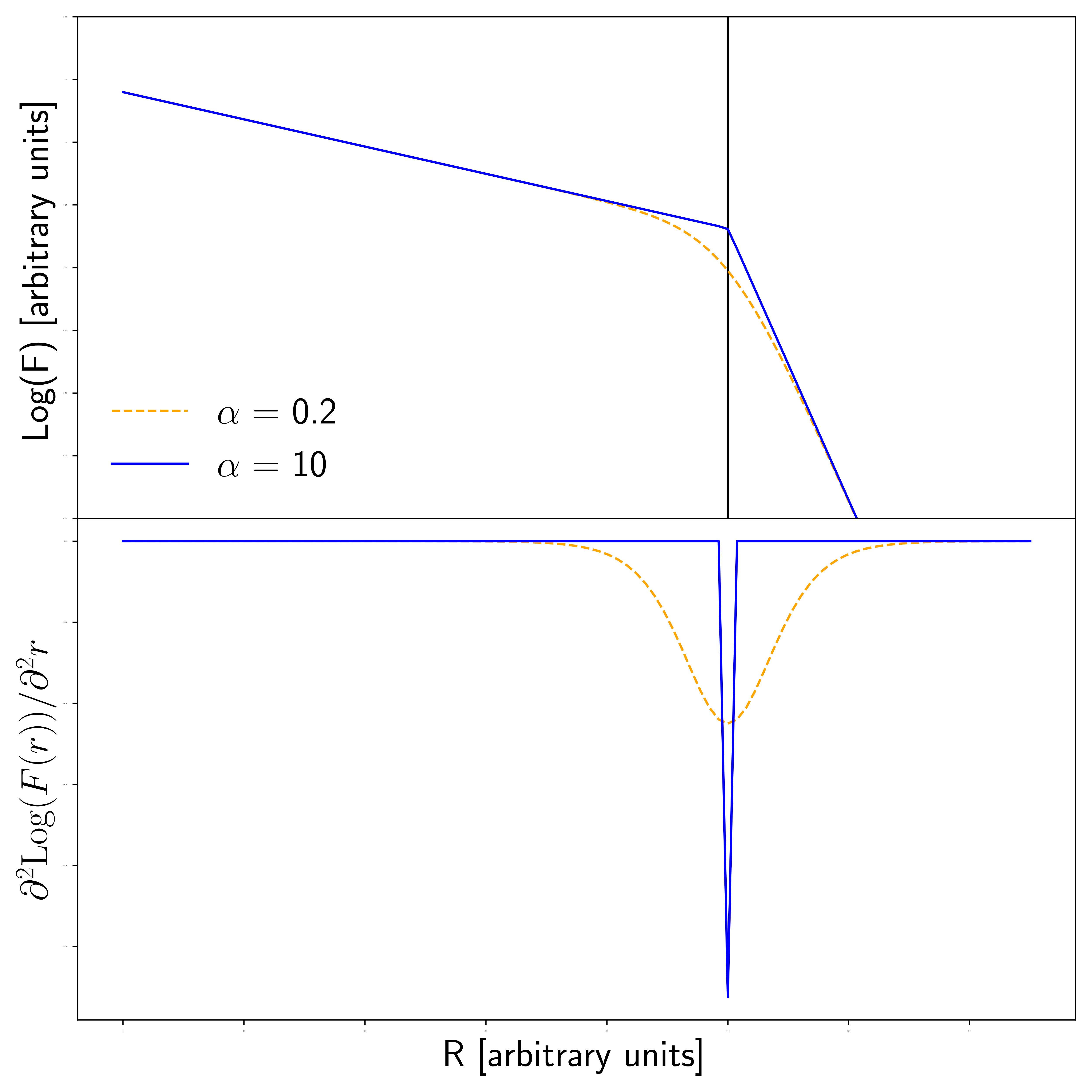}
    \caption{Derivatives of \texttt{BrokenExponential} disks with different $\alpha$ parameters.}
    \label{fig:theory}
\end{figure}

Figure \ref{fig:theory} shows the \texttt{BrokenExponential} profile in arbitrary units (top panel) and its second derivative (bottom panel) for two disks sharing the same PA, AR, central intensity, inner and outer scale length ($h_1$, $h_2$), and $R_\textnormal{edge}$ but different $\alpha$. While the location of the minimum of the second derivative remains unchanged since $R_\textnormal{edge}$ does not vary, for fixed $h_2$ a smoother transition (lower $\alpha$) leads to visually place $R_\textnormal{edge}$ at inner locations.

\section{HI isocontours}
\label{App:hi-map}

\begin{figure}
    \centering
    \includegraphics[width=\linewidth]{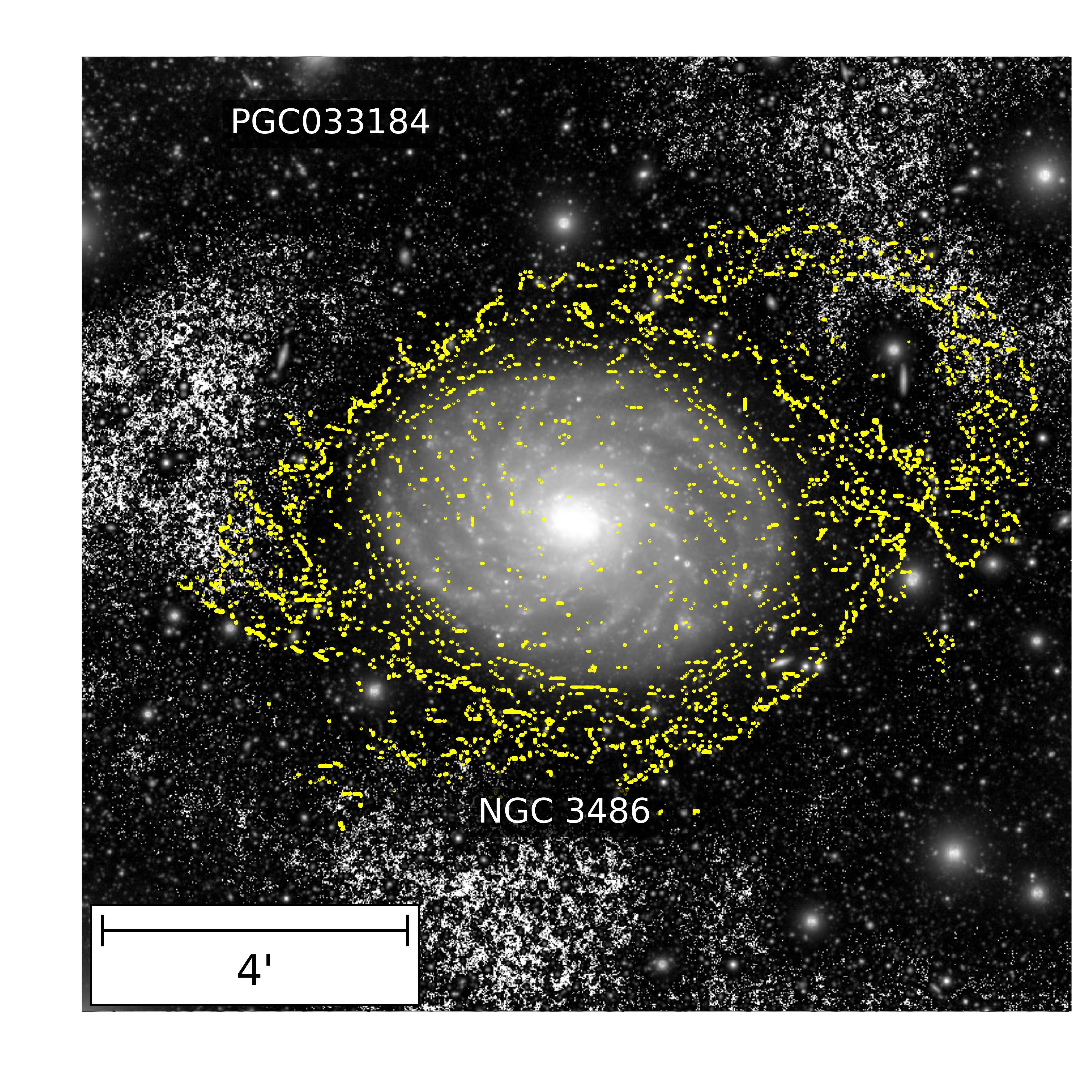}
    \caption{Low spatial resolution HI integrated intensity contours from new VLA data from \cite{2016Richardsh1} overlapped with gray scaled image of LIGHTS data of NGC 3486. }
    \label{fig:RADIOMAP-APP}
\end{figure}

Fig. \ref{fig:RADIOMAP-APP} shows the HI contours from \cite{2016Richardsh1} on top of the gray-scaled LIGHTS data of NGC 3486.

\end{appendix}

\end{document}